\documentclass[epj]{svjour}
\usepackage{graphics}

\newcommand{\re}{\mathop{\mathrm{Re}}}
\newcommand{\im}{\mathop{\mathrm{Im}}}

\begin{document}
\title{Phenomenological analysis connecting proton-proton
and antiproton-proton elastic scattering}
\author{R.F. \'Avila\inst{1}, S.D. Campos\inst{2}, M.J. Menon\inst{2}, J. Montanha\inst{2}}

\mail{menon@ifi.unicamp.br}          
\institute{Instituto de Matem\'atica, Estat\'{\i}stica e Computa\c c\~ao
Cient\'{\i}fica, Universidade Estadual de Campinas, 13083-970 Campinas, SP Brazil
\and 
Instituto de F\'{\i}sica Gleb Wataghin
Universidade Estadual de Campinas, 13083-970 Campinas, SP Brazil}
\date{Received: date / Revised version: date}
%
\abstract{Based on the behavior of the  elastic 
scattering data, 
we introduce an almost model-independent parametrization for the 
imaginary part of the scattering amplitude, with the energy 
and momentum transfer dependences inferred on empirical basis
and selected by rigorous theorems and bounds from axiomatic quantum field theory.
The corresponding real part is analytically evaluated by
means of dispersion relations, allowing connections between particle-particle and
particle-antiparticle scattering. Simultaneous fits to
proton-proton and antiproton-proton experimental data
in the forward direction and also including data beyond the forward direction, 
lead to a predictive formalism in both energy and momentum transfer.
We compare our extrapolations with predictions from some popular models
and discuss the 
applicability of the results in the normalization
of elastic rates that can be extracted from present and future accelerator experiments 
(Tevatron, RHIC and LHC).
\PACS{
      {13.85.Dz}{Elastic scattering}   \and
      {13.85.-t}{Hadron-induced high-energy interactions}
     } 
} 
\authorrunning{\'Avila, Campos, Menon, Montanha}

\titlerunning{Phenomenological analysis connecting $pp$ and $\bar{p}p$
elastic scattering}

\maketitle
\section{Introduction}
\label{sec:1}

Elastic hadron-hadron scattering, the simplest hadronic collision process,
still remains one of the topical theoretical problems in particle physics
at high energies. In the absence of a pure QCD description of these 
large-distances scattering states (soft diffraction), empirical analysis
based on model-independent fits to the physical quantities involved, play
an important role in the extraction of novel information, that can contribute with
the development of useful calculational schemes in the underlying
field theory.

In this context, empirical parametrizations of the scattering amplitude and fits to
the differential cross section data have been widely used as a source
of model-indepen\-dent determination of several quantities of interest
(the inverse problem),
such as the profile, the eikonal, the inelastic overlap functions, and, with some
additional hypothesis, even information on form factors (momentum transfer space). These
aspects were recently reviewed and discussed in \cite{cmm}, where a list of 
references to some essential results can also be found.
However, one aspect of this kind of analysis concerns its local
description of the experimental data, that is, the free parameters are inferred
from fits to each energy and to each interaction process, and therefore the approach 
has no predictive character.

In this work we present a novel parametrization for the imaginary part of the
elastic scattering amplitude with energy and momentum dependences extracted from
the empirical behavior of the experimental data and selected according to some
high-energy theorems and bounds from axiomatic quantum field theory.
The real part of the amplitude
is analytically evaluated by means of dispersion relations, connecting, therefore,
particle-particle and particle-antiparticle scattering. 
In this context, the scattering amplitude is expressed as entire functions
of the momentum transfer and of the logarithm of the energy.
Global fits
to proton-proton ($pp$) and antiproton-proton ($\bar{p}p$) experimental data 
in the forward direction
(total cross section and the $\rho$ parameter)
 and, in a second
step, also including the differential cross
sections, lead to a predictive
formalism in the energy and momentum transfer, which is also essentially model 
independent.
We present extrapolations for values of the energy and momentum transfer
above those reached in experiments and compare with predictions from some
phenomenological models. We also discuss the applicability of the results in the 
normalization of the elastic rates that can be measured in present and future
accelerator experiments (Fermilab Tevatron, Brookhaven RHIC and CERN LHC).

As will be stressed, this analysis must be seen
as a first step or attempt toward a formally rigorous model-independent description 
of high-energy
elastic hadron scattering, embodying a predictive character. In this sense we shall
attempt to
discuss and explain, in certain detail, the advantages and disadvantages of the 
present analysis and results.

The paper is organized as follows. In Sect. \ref{sec:2}
we discuss the empirical and formal bases of the parametrization for the imaginary
part of the scattering amplitude and the analytical determination of the 
corresponding real part by means of dispersion relations. In Sect. \ref{sec:3}
we present the fit procedures and results, treating firstly only the forward scattering and
in a second step,
including the differential cross section data. In Sect. \ref{sec:4} we discuss 
the physical implications and applicability of the approach, 
in the experimental and phenomenological contexts.
The conclusions and some final remarks are the contents of Sect. \ref{sec:5}.

\section{Analytical parametrization for the scattering amplitude}
\label{sec:2}

The physical quantities that characterize the elastic hadron scattering
are given in terms of the scattering amplitude $F$, which is expressed
as function of two Mandelstam variables in the center-of-mass 
system, usually the energy squared $s$ and the momentum
transfer squared $t = - q^2$.
We shall base our discussion on the following physical quantities
\cite{bc}: the differential cross section,

\begin{eqnarray}
\frac{d{\sigma}}{dq^2}=\frac{1}{16\pi s^2}
|\re F(s,q^2)+i\im F(s,q^2)|^{2}, 
\end{eqnarray}
the total cross section (optical theorem),

\begin{eqnarray}
\sigma_{tot}(s) = \frac{\im F(s, q^2 = 0)}{s}, 
\end{eqnarray}
the $\rho$ parameter (related with the phase of the amplitude
in the forward direction),

\begin{eqnarray}
\rho(s) = \frac{\re F(s,q^2 = 0)}{\im F(s,q^2 = 0)}, 
\end{eqnarray}
and the slope of the differential cross section in the forward direction,

\begin{eqnarray}
B(s) = \frac{d}{dq^2} \left[ \ln \frac{d\sigma}{dq^2}(s,q^2) \right]_{q^2=0}.
\end{eqnarray}

Equations (1) and (2) represent normalizations valid in the high-energy
region, for example $\sqrt s > $ 20 GeV \cite{bc}.
We shall return to this point in what follows.

In this Section we first discuss in certain detail the empirical and
formal bases that lead
to an almost model-independent analytical parametrization for the imaginary part of the 
amplitude in terms of both energy and momentum transfer variables. We then treat 
the analytical evaluation of the 
corresponding real part by means of derivative dispersion relations
and the analytical connections between $pp$ and $\bar{p}p$ scattering.

\subsection{Parametrization for the imaginary part of the amplitude}
 
\subsubsection{Empirical bases}

Let us first investigate some empirical information on the
differential cross section, in the
\textit{region of small momentum transfer}, 
$q^2 \leq$ 0.2 GeV$^2$. In particular, it is known that at
$q^2 = 0$, the data indicate that $|\rho(s)| \leq 0.15$, as
illustrated in Fig. 1.
Therefore, it is expected that, at small values of the momentum transfer,
the amplitude $F(s, q^2)$ to be dominantly imaginary, so that the
differential cross section in this region can be expressed as 

\begin{eqnarray}
\frac{d{\sigma}}{dq^2} \approx \frac{1}{16\pi}
\left[\frac{\im F(s,q^2)}{s}\right]^{2}. \nonumber
\end{eqnarray}

\begin{figure}
\resizebox{0.48\textwidth}{!}{\includegraphics{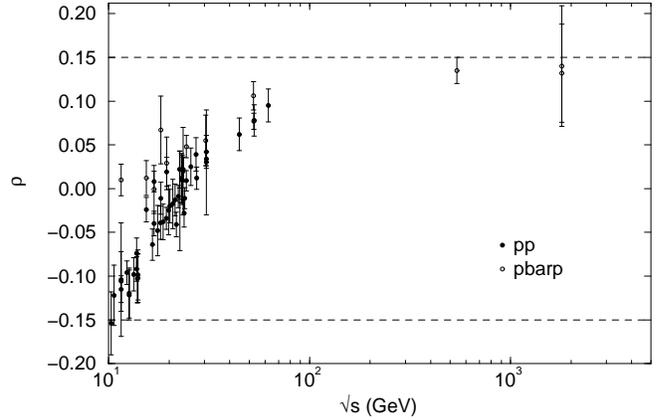}}
\caption{Dependence of $\rho$ from $pp$ and $\bar{p}p$ elastic
scattering (data from \cite{pdg,e811}).}
\label{fig:1}       
\end{figure}

Moreover, in this region, the differential cross section data are approximately
linear in the logarithm scale, as exemplified in Fig. 2, which means that we can
express the imaginary part of the amplitude as

\begin{eqnarray}
\frac{\im F(s,q^2)}{s} \approx \alpha e^{-\beta q^2}, \nonumber
\end{eqnarray}
where $\alpha$ and $\beta$ are real parameters that can depend on the
energy and reaction considered.

\begin{figure}
\resizebox{0.48\textwidth}{!}{\includegraphics{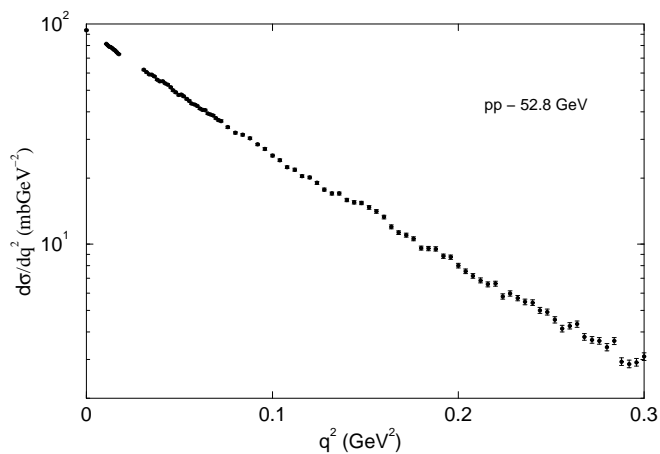}}
\caption{Diffraction peak from $pp$ elastic scattering
at $\sqrt s$ = 52.8 GeV \cite{pp53}.}
\label{fig:2}       
\end{figure}

The point now is to look for possible empirical dependences for these
parameters in terms of the energy $s$, namely analytical expressions
for $\alpha(s)$ and $\beta(s)$ and that is one of the novel aspects of
this work.
From the above two equations and from
Eqs. (2) and (4) we have that $\alpha(s) \propto \sigma_{tot}(s)$
and $\beta(s) \propto B(s)$. 
On the other hand, the empirical behavior of
$\sigma_{tot}(s)$ and  $B(s)$ (near the forward direction), displayed in Figs. 3 and 4,
respectively, indicates that
in the region of high energies 
, the empirical trends of the data (above $\sqrt s \approx$ 20 GeV)
follow polynomial dependences in $\ln s$, of second
degree (total cross section) and first degree (slope). For these
reasons it is reasonable to introduce the following 
empirical parametrizations for $\alpha(s)$ and $\beta(s)$:

\begin{eqnarray}
\alpha(s) = A + B\ln s + C \ln^2 s,
\nonumber
\end{eqnarray}
and
\begin{eqnarray}
\beta(s) = D + E\ln s, 
\nonumber
\end{eqnarray}
where $A$, $B$, $C$, $D$ and $E$ are real constants.
We note that a dimensionally necessary factor $s_0$, in $\ln s/s_0$, is
automatically absorbed by the other constants. We also note that
the choice for $\alpha(s)$ is in agreement 
with the universal asymptotic behavior of the total cross sections from the
analysis developed by the COMPETE Collaboration \cite{compete}.

\begin{figure}
\resizebox{0.48\textwidth}{!}{\includegraphics{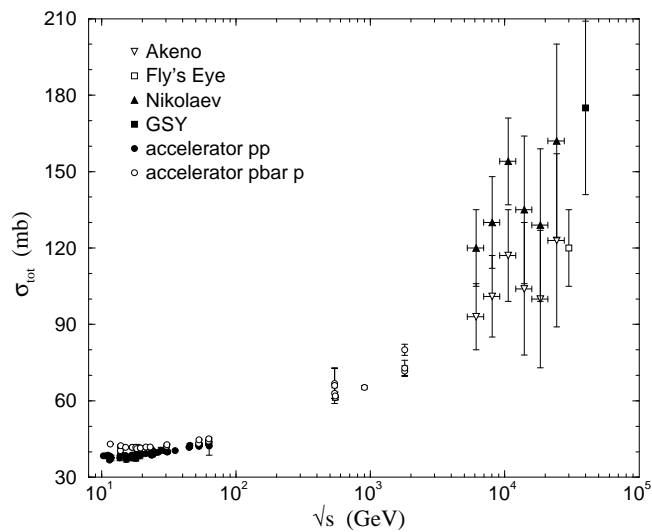}}
\caption{Experimental information on the $pp$ and
$\bar{p}p$ total cross sections from accelerator \cite{pdg} and
cosmic-ray experiments (see \cite{alm03} for a complete list of
references and discussions).}
\label{fig:3}       
\end{figure}

\begin{figure}
\resizebox{0.48\textwidth}{!}{\includegraphics{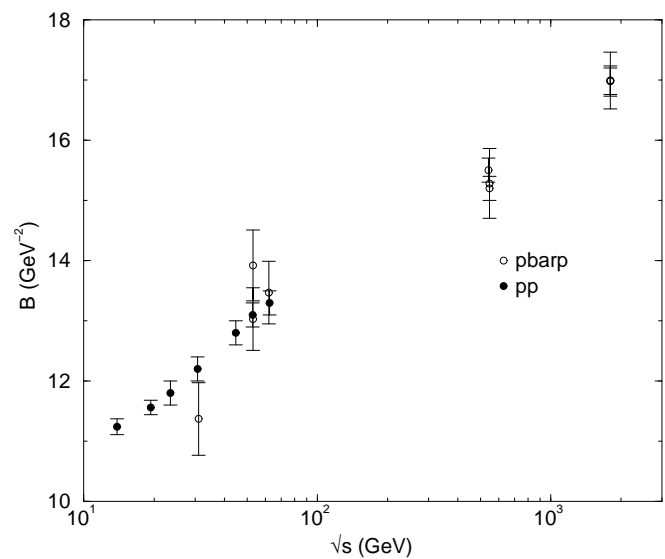}}
\caption{The slope parameter as function of the energy and determined
in the interval 0.01 $ < q^2 < $ 0.20 GeV$^2$ \cite{slope,e710}.}
\label{fig:4}       
\end{figure}

Now, in the \textit{region of medium and large momentum transfer},
the differential cross section data is characterized by the diffractive pattern, 
as illustrated in Fig. 5. Since we have a logarithmic scale, this behavior 
can be taken into account by the standard sum of exponentials in $q^2$.

\begin{figure}
\resizebox{0.48\textwidth}{!}{\includegraphics{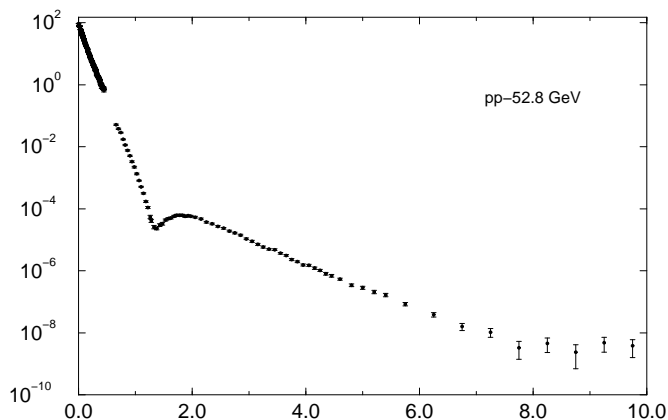}}
\caption{Proton-proton differential cross section data at $\sqrt s = 52.8$ GeV
\cite{pp53}.}
\label{fig:5}       
\end{figure}

From the above discussion and aimed to treat both
$pp$ and $\bar{p}p$ elastic scattering, we introduce the
following empirical parametrizations for $pp$ scattering,

\begin{eqnarray}
\frac{\im F_{pp}(s,q^2)}{s}={\sum_{i=1}^n}{\alpha}_i(s)
e^{-\beta_i(s)q^2},
\end{eqnarray}
with

\begin{eqnarray}
\alpha_i(s)&=&A_i+B_i\ln(s)+C_i\ln^2(s), \\ \nonumber
\beta_i(s)&=&D_i+E_i\ln(s),
\end{eqnarray}
and for $\bar{p}p$ scattering,

\begin{eqnarray}
\frac{\im F_{\bar{p}p}(s,q^2)}{s}={\sum_{i=1}^n}{\bar{\alpha}}_i(s)
e^{-\bar{\beta}_i(s)q^2},
\end{eqnarray}
with

\begin{eqnarray}
\bar{\alpha}_i(s)&=&\bar{A}_i+\bar{B}_i\ln(s)+\bar{C}_i\ln^2(s), \\ \nonumber
\bar{\beta}_i(s)&=&\bar{D}_i+\bar{E}_i\ln(s),
\end{eqnarray}
where $i=1,2,...n$. In what follows we shall check these parametrizations
in a formal context.

\subsubsection{Constraints from Axiomatic Quantum Field Theory}

Even after QCD, unitarity, analyticity, crossing and their connections
with axiomatic quantum field theory (AQFT) still remain a fundamental theoretical framework
in the investigation of high-energy \textit{soft} scattering. In this context,
important high-energy
theorems and bounds have been demonstrated \cite{eden,theo,vernov,martin},
providing rigorous formal constraints in the region of asymptotic energies,
which can not be disregarded in any reliable formalism, mainly related with
model-independent approaches.
Since parametrizations (5 - 8) were based exclusively on the behavior of the experimental
data at fixed (finite) energies, it is necessary to check the most important 
formal asymptotic results.

Firstly we note that from the optical theorem (2), the parametrizations
for $\alpha_i(s)$ and $\bar{\alpha}_i(s)$ do not violate the Froissart-Martin bound,
a rigorous prediction of quantum field theory \cite{bound}, which states that

\begin{eqnarray}
\sigma_{tot} \leq c \ln^2 s,
\nonumber
\end{eqnarray}
where $c$ is a constant.

Another important result 
concerns the behavior of the difference between particle-particle and
antiparticle-par\-ti\-cle cross sections at the asymptotic regime. In this context
it has been demonstrated by Eden \cite{eden} and  by Grunberg and Truong
\cite{gt} that if the Froissart-Martin bound is reached,
the difference between the $pp$ and $\bar{p}p$ total cross sections
goes as

\begin{eqnarray}
\Delta \sigma = \sigma_{tot}^{pp} - \sigma_{tot}^{\bar{p}p} \leq
c \frac{\sigma_{tot}^{pp} + \sigma_{tot}^{\bar{p}p}}{\ln s}, \nonumber
\end{eqnarray}
which means that the \textit{difference can increases at most as} $\ln s$ 
and even in this case,

\begin{eqnarray}
\frac{\sigma_{tot}^{\bar{p}p}}{\sigma_{tot}^{pp}} \rightarrow 1,
\nonumber
\end{eqnarray}
as $s \rightarrow \infty$ (the generalized or revised form of the
Pomeranchuk theorem).

Now, from the optical theorem, Eq. (2) and parametrizations (5-8) we have

\begin{eqnarray}
\Delta \sigma =
\sum_{i=1}^{n} \{ (A_i - \bar{A}_i) + (B_i - \bar{B}_i)\ln s + 
(C_i - \bar{C}_i) \ln^2 s \},
 \nonumber
\end{eqnarray}
and therefore, in order to not violate the above formal results, we must
impose the constraint

\begin{eqnarray}
\sum_{i=1}^{n} (C_i - \bar{C}_i) = 0.
\end{eqnarray}

With this condition we also ensure another important formal result,
namely that if the Froissart-Martin bound is reached the $\rho$ parameter
must go to zero logarithmically \cite{martincheung}

\begin{eqnarray}
\rho(s \rightarrow \infty) \propto \frac{1}{\ln s}.
\nonumber
\end{eqnarray}

With parametrizations (5-8) and the constraint (9) we have $10 n - 1$ free 
parameters, where $n$ is the number
of exponentials. The
novelty in these parametrizations is the fact that the energy dependences 
are already enclosed and were inferred from the empirical behavior of the
experimental data, being also in agreement with the above high-energy theorems.
Moreover, the imaginary parts of the amplitudes are entire functions of the
logarithm of the energy, which is an important property in the evaluation
of the real part, as discussed in what follows.

\subsection{Analytical evaluation of the real part of the amplitude}

Connections between real and imaginary parts of the \textit{forward}
scattering amplitude have been widely investigated by means of dispersion
relations in both \textit{integral and derivative forms}. In this work we make use
of the \textit{derivative relations} \cite{ddr,bks}, which are valid in the 
forward direction and
for amplitudes belonging to a sub-class of 
entire functions of the logarithm of the energy, as it is our case.
For a recent review and critical
analysis on the replacement of integral relations by derivative ones see Ref. 
\cite{am04}, where a list of references to some outstanding works can also be found. 

In the \textit{forward direction}, the derivative dispersion relations for even ($+$) and odd
($-$) amplitudes are expressed in terms of a tangent operator and in the
case of one subtraction (equal to two subtractions in the even case) they
are given by \cite{ddr,bks,am04}

\begin{equation}
\frac{\re F_+(s)}{s}= \frac{K}{s} +
\tan\left[\frac{\pi}{2}
\frac{\mathrm{d}}{\mathrm{d}\ln s}\right]\frac{\im F_+(s)}{s},
\end{equation}

\begin{equation}
\frac{\re F_-(s)}{s}=
\tan\left[\frac{\pi}{2}
\left(1 + \frac{\mathrm{d}}{\mathrm{d}\ln s}\right)\right]
\frac{\im F_-(s)}{s},
\label{eq:11}
\end{equation}
where $K$ is the subtraction constant. It has also been demonstrated 
by Fischer and Kol\'a\v{r} \cite{kf} that at high energies the above tangent operator
can be replaced by its first order expansion, which is the case we
are interested in.

Since besides the forward data we are also aimed to investigate differential
cross sections, it is necessary to consider the applicability
of the dispersion techniques beyond the forward direction. Although several authors
make use of dispersion relations even for large values of the momentum
transfer, it is important
to recall that what is formally expected is the validity of
the dispersion relations, with a finite number of subtractions, inside
a region $q^2 \leq q_{max}^2$.
However, the exact expression and/or numerical value of $q_{max}^2$ depends on 
the theoretical framework and scattering process considered. We shall discuss this
subject in some detail in Sect. 3.2.2, when applying the formalism to the experimental
data. Here we only consider a reference to this limited interval.

Based on the above arguments we shall make use of the \textit{first order} derivative
dispersion relations, also extended beyond the forward region, namely
$0 \leq q^2 \leq q_{max}^2$, in the form

\begin{eqnarray}
\frac{\re F_+(s, q^2)}{s}=  \frac{K}{s} +
\frac{\pi}{2}
\frac{\mathrm{d}}{\mathrm{d}\ln s}\frac{\im F_+(s, q^2)}{s},
\end{eqnarray}

\begin{eqnarray}
\frac{\re F_-(s, q^2)}{s}=
\frac{\pi}{2}
\left(1 + \frac{\mathrm{d}}{\mathrm{d}\ln s}\right)
\frac{\im F_-(s, q^2)}{s}.
\end{eqnarray}

Finally, the connections between the hadronic and the even/odd amplitudes
is established through the usual definitions:

\begin{eqnarray}
F_{pp}(s, q^2) = F_{+}(s,q^2) + F_{-}(s,q^2) \\ \nonumber
F_{\bar{p}p}(s, q^2) = F_{+}(s,q^2) - F_{-}(s,q^2)
\end{eqnarray}

This approach is characterized by analytical results for both real
and imaginary parts of the $pp$ and $\bar{p}p$ amplitudes.
Schematically, from parametrizations (5 - 8) for
$\im F_{pp/\bar{p}p}(s, q^2)/s$
we obtain
$\im F_{+/-}(s, q^2)/s$ by inverting Eqs. (14). Then the
derivative relations (12 - 13) allow to  evaluate 
$\re F_{+/-}(s, q^2)/s$ and by Eqs. (14) we obtain the hadronic real parts

\begin{eqnarray}
\frac{\re F_{pp}(s,q^2)}{s}&=& \frac{K}{s}  \nonumber \\
&+&\sum_{i=1}^n \left\{
\frac{\pi}{2}\left[\alpha_i'(s)-\alpha_i(s)\beta_i'(s)q^2\right]
e^{-\beta_i(s) q^2} \right. \nonumber \\ 
 &+&\left.\frac{\pi}{4}
\left[\alpha_i(s) e^{-\beta_i(s) q^2}-\bar{\alpha_i}(s)e^{-\bar{\beta_i}(s) q^2}\right]
\right\}, \nonumber
\end{eqnarray}

\begin{eqnarray}
\frac{\re F_{\bar{p}p}(s,q^2)}{s}&=& \frac{K}{s}  \nonumber \\
&+&\sum_{i=1}^n\left\{
\frac{\pi}{2}\left[\bar{\alpha_i}'(s)-\bar{\alpha_i}(s)\bar{\beta_i}'(s)q^2\right]
e^{-\bar{\beta_i}(s) q^2} \right. \nonumber \\
&-&\left.\frac{\pi}{4}
\left[\alpha_i(s) e^{-\beta_i(s) q^2}-\bar{\alpha_i}(s)e^{-\bar{\beta_i}(s) q^2}\right]
\right\}, \nonumber
\end{eqnarray}
where the primes denote differentiation with respect to $\ln s$
(Eqs. (6) and (8)).
With this we have analytical
expressions for the
$pp$ and $\bar{p}p$ differential cross sections:

\begin{eqnarray}
\frac{d\sigma_{pp/\bar{p}p}}{dq^2}=
\frac{1}{16\pi}
\left|\frac{\re F_{pp/\bar{p}p}(s,q^2)}{s}+
i \frac{\im F_{pp/\bar{p}p}(s,q^2)}{s}\right|^{2}. \nonumber
\end{eqnarray}

It should be noted that exact analyticity and crossing properties demand
symmetric variables, namely the laboratory energy $E$ for $q^2$ = 0 and
the variable $(s - u) /4m$, where $u$ is the Mandelstam variable, for
$q^2 >$ 0 \cite{bc}. However, since $E$ depends linearly on $s$ and, as 
will be discussed 
in Sect. 3.2, we shall consider the applicability of the formalism mainly
in limited regions of the momentum transfer and only above $\sqrt s$ = 20 GeV, 
the use of $s$ as variable does not introduce essential changes in the
above formulas.

Taking into account the subtraction constant $K$, the constraint (9) and
parametrizations (5-8) we eventually have 10$n$ fit parameters in the case
of $n$ exponential terms. 
This completes the analytical construction of the formalism, characterized
by its empirical basis, essentially model-independent parametrizations,
agreement with high-energy theorems
 and amplitudes belonging to the class of entire
functions in the logarithm of the energy. In the next
section we determine the free parameters involved through fits to
$pp$ and $\bar{p}p$ elastic scattering data.

\section{Experimental data, fitting and results}
\label{sec:3}

\subsection{Experimental Data}

The most important empirical input in our parametrization is the
energy dependence enclosed in the expressions of $\alpha(s)$ and $\beta(s)$,
Eqs. (6) and (8), respectively. Since it characterizes the region where
the \textit{total cross section increases with the energy}, we shall consider 
here only
the experimental data available above $\sqrt s$ = 20 GeV
from $pp$ and $\bar{p}p$ scattering. 
We note that this necessary threshold puts limitations
in extensions of the formalism to other reactions, such as 
$\pi^{\pm}p$, $K^{\pm}p$, etc..., due to the small number of experimental data
available.

For the forward data on $\sigma_{tot}$ and $\rho$, we use the Particle Data
Group archives \cite{pdg}, to which we added the value of $\rho$ and $\sigma_{tot}$ 
at 1.8 TeV obtained
by the E811 Collaboration \cite{e811}. The statistical and systematic errors were 
added in quadrature. We did not include the cosmic-ray information on $pp$
total cross sections due to the model dependences involved \cite{alm03}.

The differential cross section data include the optical point,

\begin{eqnarray}
\left[\frac{d\sigma(s,q^2)}{dq^2}\right]_{q^2 = 0} =
\frac{\sigma_{tot}(1+\rho^2)}{16\pi},
\end{eqnarray}
and the data above the Coulomb-nuclear interference region,
namely $q^2 > 0.01$ GeV$^2$. The data include 12 sets form
$pp$ scattering, at $\sqrt s$ = 23.5, 27.4, 30.7, 44.7, 52.8, and 62.5 GeV and 
from $\bar{p}p$ scattering, at $\sqrt s$ =  31, 53, 61, 546, 630 and 1800 GeV. 
The  $pp$ data
at 27.4 GeV, covering the region 
5.5 $\leq q^2 \leq $ 14 GeV$^2$, are from \cite{faissler}.
The $\bar{p}p$ differential cross section data at 1.8 TeV
include those obtained by the E710 Collaboration \cite{e710} 
(0.045 $\leq q^2 \leq $ 0.627 GeV$^2$)
and by
the CDF Collaboration \cite{cdfdcs} (0.035 $\leq q^2 \leq $ 0.285 GeV$^2$).
In this case we used two optical points with the values of $\sigma_{tot}$
and $\rho$ from references \cite{e811} (E811 Collaboration) and \cite{e710}
(E710 Collaboration).
The complete list of references to the other
data sets can be found in \cite{cmm} (references [26, 28-31]).
In all these sets the experimental errors
correspond to the statistical ones.

We note that we have used all the experimental data referred to before, 
that is, we did not performed any kind of data selection in the above
standard ensemble.

\subsection{Fitting and results}

As recalled in Sec. 2.2, the applicability of the dispersion relations,
outside the forward direction, depends on the maximum value of the momentum
transfer considered. For this reason, we shall treat separately fits to only
forward quantities, $\sigma_{tot}$ and $\rho$, and simultaneous
fits to these quantities plus the differential cross section data.
We first present the fits to the forward data and next
discuss the applicability of the dispersion techniques
beyond the forward direction. 

\subsubsection{Fits to the forward scattering data}

Making use of the formalism described in Sect. 2, we 
performed simultaneous fits to $\sigma_{tot}$ and $\rho$ data, above 20 GeV,
from $pp$ and $\bar{p}p$ scattering. Since we are treating here only forward
data ($q^2$ = 0), the sole parameters involved are those associated with
$\alpha_i(s)$ and $\bar{\alpha}_i(s)$ in Eqs. (6) and (8).

The fits were performed through the CERN-Minuit code, with the
estimated errors in the free parameters corresponding to an increase of the
$\chi^2$ by one unit. For this ensemble of data good statistical results were
obtained with only one exponential factor ($n=1$ in Eqs. (5) and (7)) and the
best fit indicated $\chi^2$/DOF =1.07 for 83 degrees of freedom. The constraint (9)
in this case reduces to $C_1 = \bar{C}_1$ and the fit indicated a value of the
subtraction constant compatible with zero.
The numerical results are displayed in Table 1 and the corresponding curves,
together with the experimental data analyzed, are shown in Figure 6.

\begin{table*}
\begin{center}
\caption{Results of the simultaneous fits to $\sigma_{tot}$
and $\rho$ from $pp$ and $\bar{p}p$ scattering. All the parameters
are in GeV$^{-2}$,
$\chi^2$/DOF =1.07 for 83 degrees of freedom and $C_1 = \bar{C}_1 $.}
\label{tab:1}
\begin{tabular}{ccccc}

\hline\noalign{\smallskip}

\multicolumn{2}{c}{$pp$  scattering} &  \multicolumn{2}{c}{$\bar{p}p$
scattering} \\

\noalign{\smallskip}\hline\noalign{\smallskip}

$A_1$ & 121.9 $\pm$ 2.7  & $\bar{A}_1$ & 140.8  $\pm$ 3.6 \\

$B_1$ &  -9.82 $\pm$ 0.72  & $\bar{B}_1$ &  -11.78 $\pm$ 0.86   \\

$C_1$ &  1.036 $\pm$ 0.049 & $\bar{C}_1$ & 1.036 $\pm$ 0.049  \\

\hline
\end{tabular}
\end{center}
\end{table*}

\begin{figure}
\resizebox{0.48\textwidth}{!}{\includegraphics{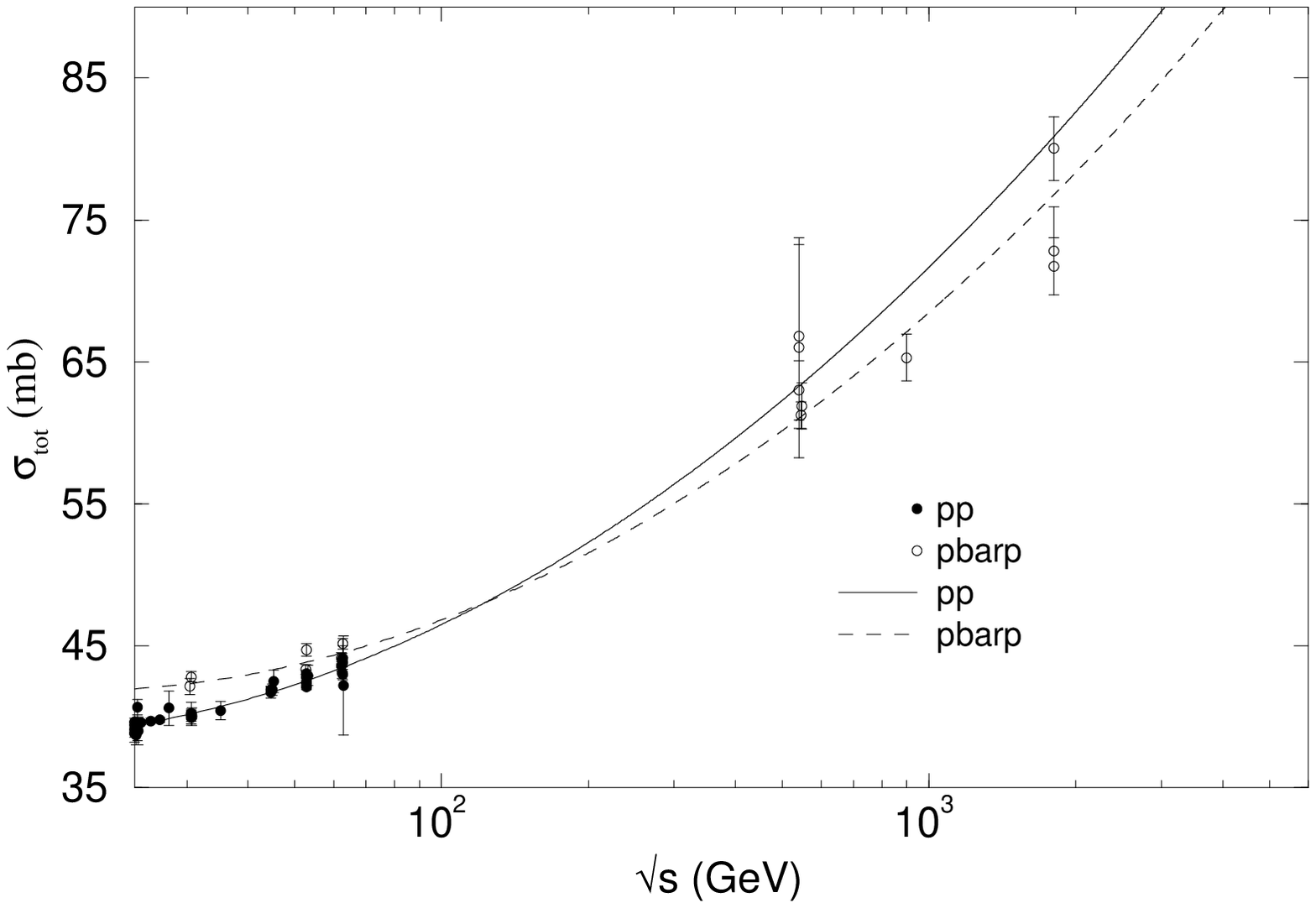}}
\resizebox{0.48\textwidth}{!}{\includegraphics{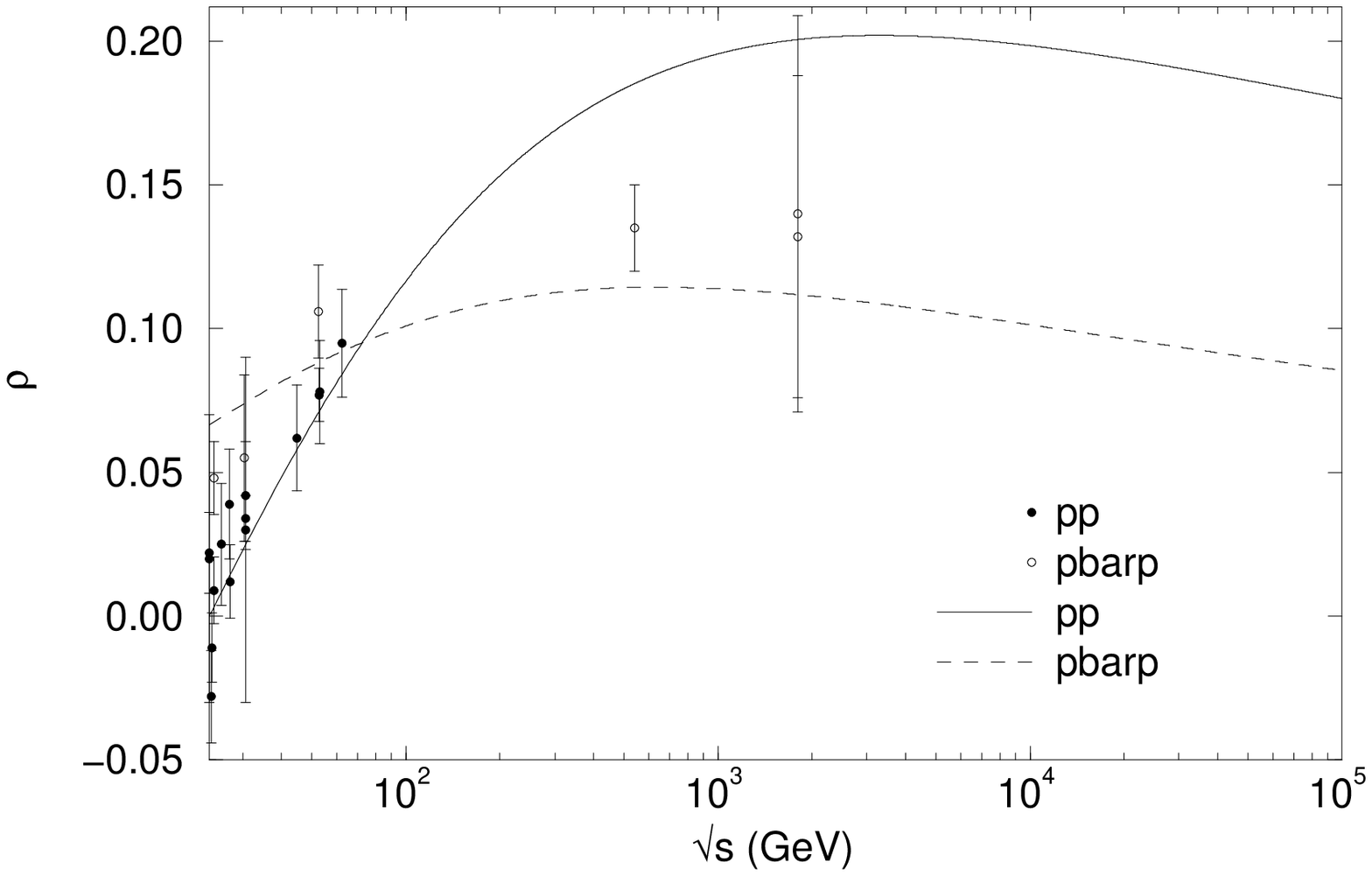}}
\caption{Fit results to the forward data, $\sigma_{tot}$
and $\rho$, from $pp$ and $\bar{p}p$ scattering.}
\label{fig:6}       
\end{figure}

These results will be discussed in detail in Sect. 4, but we note
here the good quality of the fit in terms of the $\chi^2/DOF$ and also the small number
of free parameters involved: 5. We also note a 
crossing in the total cross sections, with
$\sigma_{tot}^{pp}$ becoming higher than $\sigma_{tot}^{\bar{p}p}$
above $\sqrt s \approx $ 100 GeV, and a similar effect is predicted for $\rho(s)$. As 
we recalled in Sect. 2.1, these behaviors do not violate any high-energy
theorem on elastic hadron scattering. However, the result for $\rho^{\bar{p}p}(s)$
is below the experimental data available at the highest energies. We shall discuss
this effect in Sec. 4.2.

\subsubsection{Fits beyond the forward direction}

We now consider simultaneous fits to $\sigma_{tot}(s)$, $\rho(s)$ and 
$d\sigma(s,q^2)/dq^2$, from $pp$ and $\bar{p}p$ scattering. As recalled before,
although 
dispersion relations have been used even in the region of medium and large momentum
transfer (see, for example, \cite{bks,blp,ggk,kroll,erasmo}), 
an important point concerns the exact region in the $q^2$ variable inside
which dispersion relations hold. In what follows we first review some formal
results involved which show us that, in the case of $pp$ and $\bar{p}p$
scattering (and nucleon-nucleon in general), the situation is not simple
or neat. Based on these results we shall infer a reasonable strategy, not proved
to be wrong, that will allow us to develop simultaneous fit procedures including
the differential cross section data. Note that, since we are treating with a
sub-class of entire 
functions in the logarithm of the energy, the discussion that follows applies equally well
to both integral and derivative dispersion relations \cite{am04}.

\vspace{0.7cm}

\noindent
$\bullet$ \textit{Analyticity in} $q^2$

Dispersion relations are connected with the unitarity and analyticity
properties of the amplitude. Recently, the axiomatic approach to high-energy 
hadron scattering, as well as the rigorous analyticity-unitarity program
have been nicely reviewed in the excellent papers by
Vernov and Mnatsakanova \cite{vernov} and Martin \cite{martin},
where a complete list of references and credits to outstanding results and
authors can be found. For this reason, based on these works, we shall only summarize 
and quote here
the results, we understand, give an updated view on the $q^2$-interval inside which
dispersion relations hold. The results are the followings.

\begin{enumerate}

\item For meson-meson and meson-nucleon scattering, rigorous formalism 
based on local field theory, allows to prove that dispersion relations are valid
in finite intervals of the momentum transfer. For example,
$q_{max}^2 = 32 m_{\pi}^2/3 \approx$ 0.2 GeV$^2$ in the case of $\pi N$ scattering
\cite{vernov} and $q_{max}^2 = 28 m_{\pi}^2 \approx$ 0.55 GeV$^2$ in the case 
of $\pi \pi$ scattering \cite{martin}.

\item Similar intervals can be deduced for other processes, like
$\gamma + N \rightarrow \gamma + N^{(*)}$, $\gamma + N \rightarrow \pi + N^{(*)}$,
$e + N \rightarrow e + \pi + N^{(*)}$ \cite{martin}.

\item In the formal analyticity-unitarity context there seems to be no results
for nucleon-nucleon scattering. However, a limit
$q_{max}^2 = m_{\pi}^2/4  \approx$ 0.005 GeV$^2$ can be inferred
from perturbation theory
\cite{martin}.

\item The reason why elastic $pp$ and $\bar{p}p$ amplitudes
``lack the usual analytical properties is that the
cut
in the complex $s$ plane starts from $s_0$ = 4m$^2$ (due to virtual annihilation
process), while the physical region of $\bar{p}p$ scattering starts from $s_1$ =
4M$^2$'' \cite{vernov}. Here $m$ corresponds to the pion mass and
$M$ to the proton mass.

\item In the context of the double-dis\-per\-sion representation
by Mandelstam \cite{mandelstam1}, the domain in the $q^2$ variable, inside
which dispersion relations hold for a process $m + m \rightarrow m + m$,
extends up to $q_{max}^2 = 9m^2$ \cite{mandelstam2}. Although the
original approach treated only pion-pion scattering, one point to stress is the
fact that for all mass cases this representation was never proved nor
disproved in the contexts of the axiomatic field theory
or perturbation theory \cite{martin}. 
\end{enumerate}

We also recall that 
fixed-$q^2$ dispersion relations for nucleon-nucleon scattering have been
used by Kroll and co-authors \cite{ggk,kroll} and in particular, in Ref. \cite{ggk}, 
$pp$ and 
$\bar{p}p$ scattering were investigated through dispersion relations
in the region $0 \leq q^2 \leq 3$ GeV$^2$. However, there is no reference to a formal 
numerical value for $q_{max}^2$.

These are the results we have found and compiled on the applicability of dispersion
relations beyond the forward direction. We understand that the quoted bound from
perturbation theory seems unreliable to be considered in the case of a soft
processes like \textit{elastic} $pp$ and $\bar{p}p$ scattering. A second aspect
concerns the fact that, in the context of the axiomatic field theory, the
Mandesltam representation for all mass cases was never proved to be incorrect
(or correct either). If the representation can be extended to the
$pp$ and $\bar{p}p$ case, the analyticity domain could cover the region up to
$q_{max}^2 = 9m_{p}^2 \approx 8$ GeV$^2$.

\vspace{0.5cm}

$\bullet$ \textit{Strategies and fits}

Based on the above information,
we understand that it can be instructive to perform tests on distinct 
values for $q_{max}^2$ and investigate the
consequences in the description of the bulk of the experimental data
on $\sigma_{tot}(s)$, $\rho(s)$ and 
$d\sigma(s,q^2)/dq^2$, from $pp$ and $\bar{p}p$ scattering.
Since the typical mass scale in the hadronic scattering is the 
proton mass
(which is also expected to represent an interface between soft and semihard
processes),
it may be reasonable and perhaps even conservative, to consider some
bounds $q_{max}^2$
inside the region 1 - 2 GeV$^2$.
Moreover, it seems also important to address the practical applicability
of the dispersion approach at medium and large values of the momentum
transfer by taking into account all the
differential cross section data available, namely,
$q_{\mathrm{max}}^2$ = 14 GeV$^2$. Despite the lack of a formal justification
for this extreme case, we understand that it may also be 
useful to get some additional information on the regions where dispersion
relations work, even if only in an strictly phenomenological context.
It is important to stress that the strategy to consider different values
for $q_{\mathrm{max}}^2$ is only an \textit{ansatz} and that the main point in favor 
of this hypothesis is the
fact that there is no formal proof against it, or in other words, we understand
that it should not constitute a serious formal drawback.

Based on the above discussion,
we shall consider four variants for the fits by selecting
differential cross section data up to 
$q_{\mathrm{max}}^2$ = 1.0, 1.5, 2.0 and 14 GeV$^2$.
As before, the fits were performed through the Minuit code. For these
ensemble of data, independently of the value considered for $q_{\mathrm{max}}^2$,
 the best results demanded three exponential terms in the imaginary
part of the amplitude and therefore 30 free parameters to be fitted. The constraint
(9) was taken into account by defining $C_1=\bar{C_1}+\bar{C_2}+\bar{C_3}-C_2-C_3$.

The $\chi^2$ information on each of the four variants considered is displayed in Table 2.
We note that the $\chi^2$/DOF lies in the interval 2.5 - 3.0 for a number
of degrees of freedom equal
or greater than 923.
It is important to mention that these values are typical of global fits to the 
experimental data on
$\sigma_{tot}(s)$, $\rho(s)$ and 
$d\sigma(s,q^2)/dq^2$, from $pp$ and $\bar{p}p$ scattering \cite{dgp}.
The ``large" values are consequences of several points in the differential cross
sections that lies outside a normal distribution, as well as different normalizations
from different experiments in distinct kinematic intervals. As commented
before we did not perform any kind of data selection.

\begin{table*}
\begin{center}
\caption{Statistical information on the fit results to $\sigma_{tot}$,
$\rho$ and $d\sigma/dq^2$ data from $pp$ and $\bar{p}p$ scattering
in different intervals of the momentum transfer variable.}
\label{tab:2}
\begin{tabular}{llll}
\hline\noalign{\smallskip}
$q_{\mathrm{max}}^2$ GeV$^2$ & NDOF & $\chi^2$/DOF \\
\noalign{\smallskip}\hline\noalign{\smallskip}
1.0 & 923  & 2.476 \\
1.5 & 1003 & 2.909 \\
2.0 & 1064 & 2.881 \\
14.0 (all data) & 1277 & 2.829 \\
\noalign{\smallskip}\hline
\end{tabular}
\end{center}
\end{table*}

In fact, despite the large values of the $\chi^2$/DOF,
the visual description of the experimental data is  good in all
the cases investigated. In particular we display here the results for 
$q_{\mathrm{max}}^2$ = 2 GeV$^2$, which 
we understand can be considered a conservative case (in agreement with the
expected analyticity interval in terms of the momentum
transfer) and for $q_{\mathrm{max}}^2$ = 14 GeV$^2$.
The values of the free parameters in both cases are shown in Table 3
and the corresponding curves together with the experimental data analyzed in Figs.
7 and 8 ($q_{\mathrm{max}}^2$ = 2 GeV$^2$) and 9 and 10
($q_{\mathrm{max}}^2$ = 14 GeV$^2$). We see that the description of all the
differential cross section data is quite good, even in the extreme case
$q_{\mathrm{max}}^2$ = 14 GeV$^2$.

\begin{table*}
\begin{center}
\caption{Results of the simultaneous fits to $\sigma_{tot}$,
$\rho$ and $d\sigma/dq^2$ from $pp$ and $\bar{p}p$ scattering,
with differential cross section data up to
$q_{max}^2$ = 2.0 GeV$^2$ and $q_{max}^2$ = 14 GeV$^2$,
for which $K = 49.7 \pm 1.7$ and $K = -0.1053 \pm 0.0048$,
respectively. All the parameters are in GeV$^{-2}$ and
$C_1=\bar{C_1}+\bar{C_2}+\bar{C_3}-C_2-C_3$ .}
\label{tab:3}
\begin{tabular}{ccccccc}
\hline\noalign{\smallskip}

 \multicolumn{3}{c}{$pp$  scattering}  & \multicolumn{3}{c}{$\bar{p}p$
 scattering}  \\

\noalign{\smallskip}\hline\noalign{\smallskip}

 &$q_{max}^2$ = 2.0 GeV$^2$&$q_{max}^2$ = 14 GeV$^2$& &
$q_{max}^2$ = 2.0 GeV$^2$&$q_{max}^2$ = 14 GeV$^2$ \\

\noalign{\smallskip}\hline\noalign{\smallskip}

$A_1$ & 91.13      $\pm$ 0.28       & 109.70  $\pm$ 0.28    & $\bar{A}_1$ & 119.61   $\pm$ 0.44    & 112.28   $\pm$ 0.44 \\
$B_1$ & -12.939    $\pm$ 0.039      & -16.529 $\pm$ 0.039   & $\bar{B}_1$ & -2.486   $\pm$ 0.073   & -0.468   $\pm$ 0.074 \\
$C_1$ &  constrained  &  constrained                        & $\bar{C}_1$ & -0.0174  $\pm$ 0.0038  & -0.1673  $\pm$ 0.0039 \\
$D_1$ & -7.79      $\pm$ 0.33       & -8.91   $\pm$ 0.32    & $\bar{D}_1$ & 3.134    $\pm$ 0.067   & 3.170    $\pm$ 0.069 \\
$E_1$ & 2.908      $\pm$ 0.051      & 3.045   $\pm$ 0.050   & $\bar{E}_1$ & 0.4884   $\pm$ 0.0078  & 0.4860   $\pm$ 0.0082 \\
$A_2$ & 16.82      $\pm$ 0.22       & -4.06   $\pm$ 0.23    & $\bar{A}_2$ & 14.51    $\pm$ 0.15    & 10.23    $\pm$ 0.15 \\
$B_2$ & 7.071      $\pm$ 0.030      & 11.387  $\pm$ 0.030   & $\bar{B}_2$ & -6.730   $\pm$ 0.024   & -6.756   $\pm$ 0.027 \\
$C_2$ & 0.3027     $\pm$ 0.0047     & 0.0952  $\pm$ 0.0047  & $\bar{C}_2$ & 0.9035   $\pm$ 0.0027  & 0.9613   $\pm$ 0.0029 \\
$D_2$ & 1.647      $\pm$ 0.014      & 1.290   $\pm$ 0.014   & $\bar{D}_2$ & -1.549   $\pm$ 0.011   & -1.476   $\pm$ 0.013 \\
$E_2$ & 0.4646     $\pm$ 0.0022     & 0.5097  $\pm$ 0.0023  & $\bar{E}_2$ & 0.5521   $\pm$ 0.0011  & 0.5645   $\pm$ 0.0012 \\
$A_3$ & 0.2582     $\pm$ 0.0049     & 1.0554  $\pm$ 0.0077  & $\bar{A}_3$ & -17.160  $\pm$ 0.055   & -8.148   $\pm$ 0.031 \\
$B_3$ & -0.09894   $\pm$ 0.00074    & -0.3607 $\pm$ 0.0013  & $\bar{B}_3$ & 2.6083   $\pm$ 0.0060  & 1.2313   $\pm$ 0.0040 \\
$C_3$ & 0.5921E-02 $\pm$ 0.0070E-02 & 0.02372 $\pm$ 0.00012 & $\bar{C}_3$ & -0.11298 $\pm$ 0.00038 & -0.05572 $\pm$ 0.00025 \\
$D_3$ & 0.303      $\pm$ 0.019      & 0.6454  $\pm$ 0.0081  & $\bar{D}_3$ & 1.5672   $\pm$ 0.0084  & 0.9272   $\pm$ 0.0072 \\
$E_3$ & 0                           & 0.0176  $\pm$ 0.0011  & $\bar{E}_3$ & 0                      & 0.03859  $\pm$ 0.0011 \\

\noalign{\smallskip}
\hline
\end{tabular}
\end{center}
\end{table*}

\begin{figure}
\resizebox{0.48\textwidth}{!}{\includegraphics{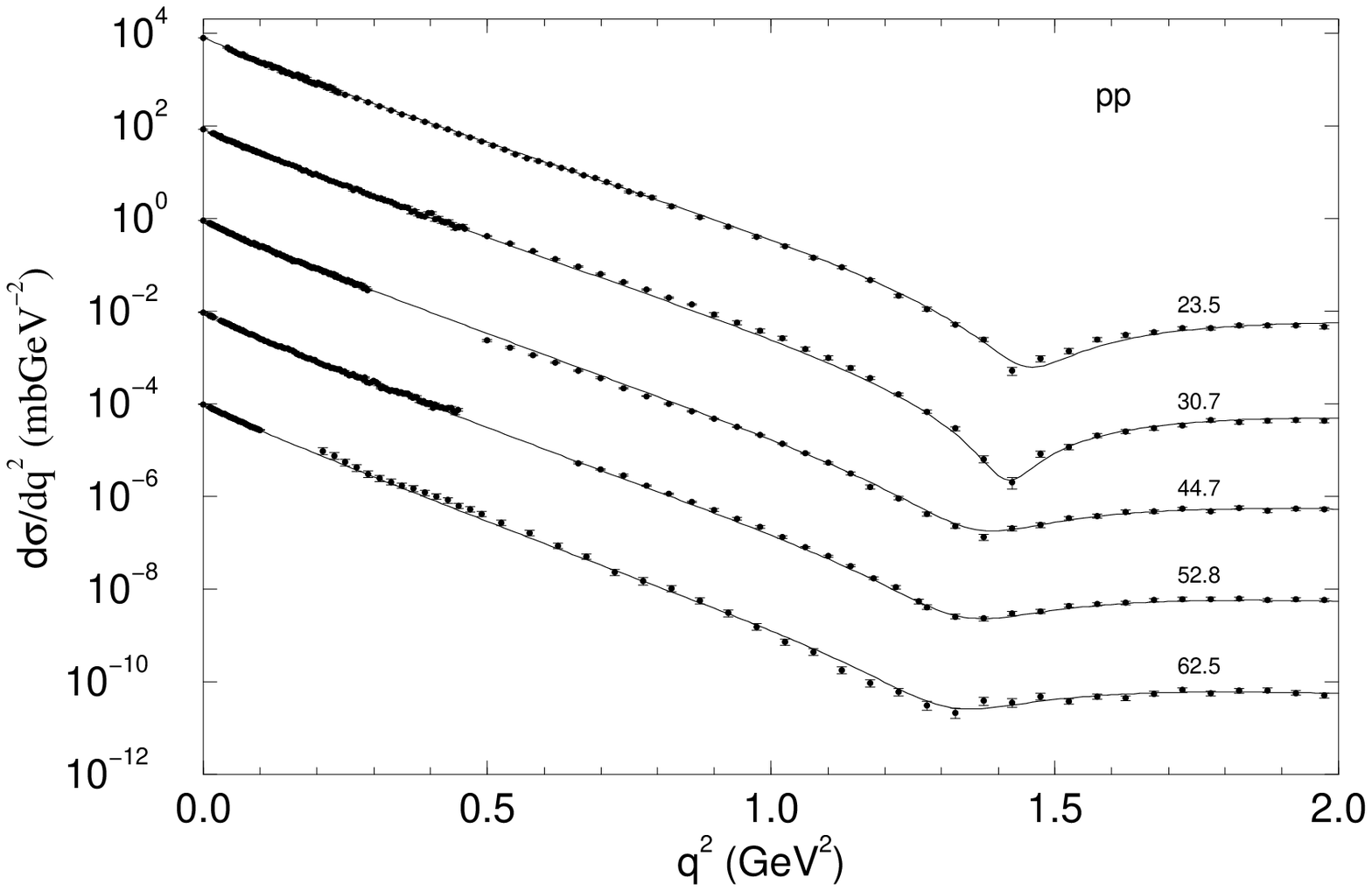}}
\resizebox{0.48\textwidth}{!}{\includegraphics{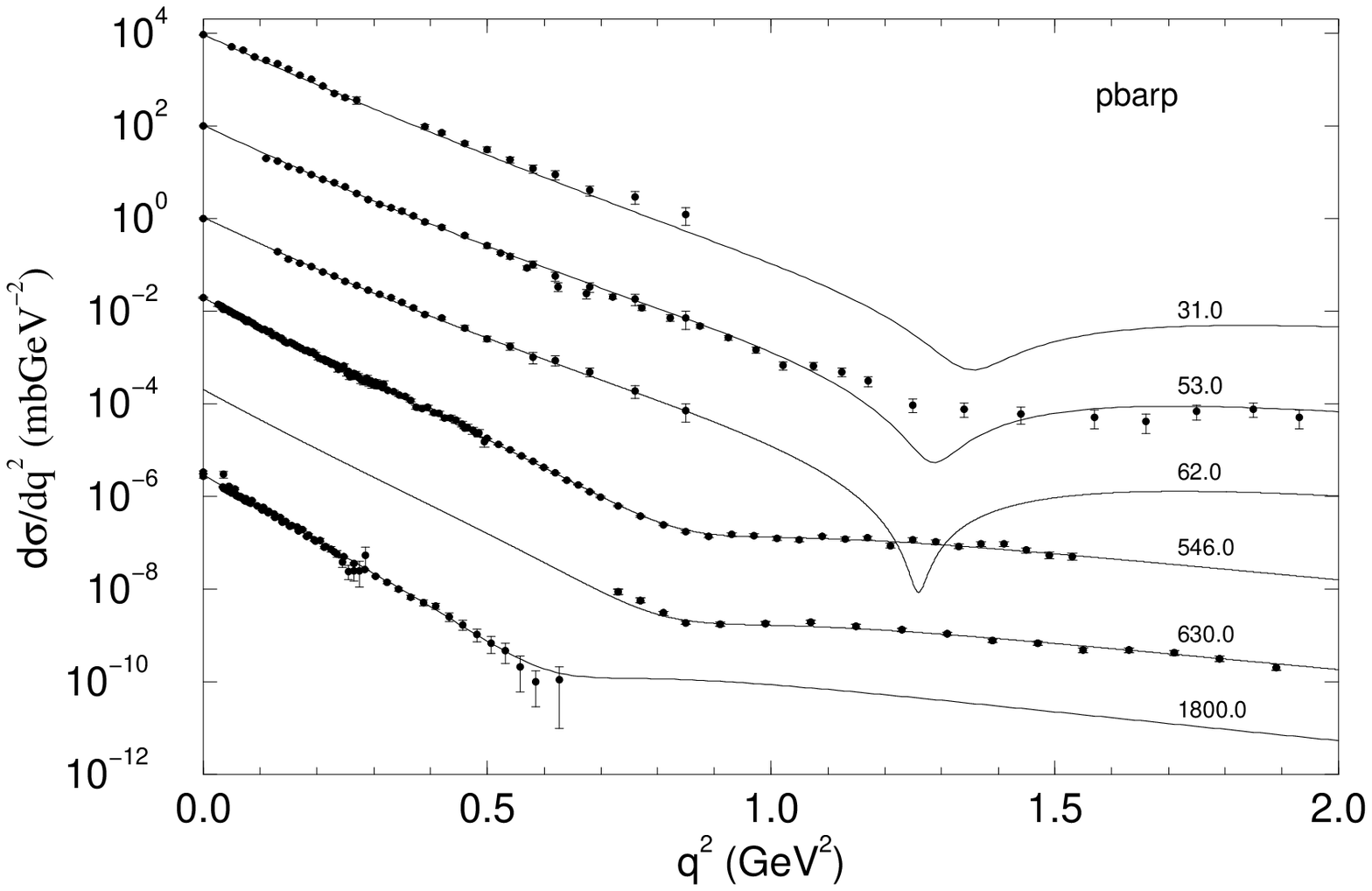}}
\caption{Differential cross sections from global fits to $pp$ 
and $\bar{p}p$ data 
with 
$q_{\mathrm{max}}^2$ = 2 GeV$^2$. Curves and data were multiplied by 
factors
of $10^{\pm 2}$.}
\label{fig:7}       
\end{figure}

\begin{figure}
\resizebox{0.48\textwidth}{!}{\includegraphics{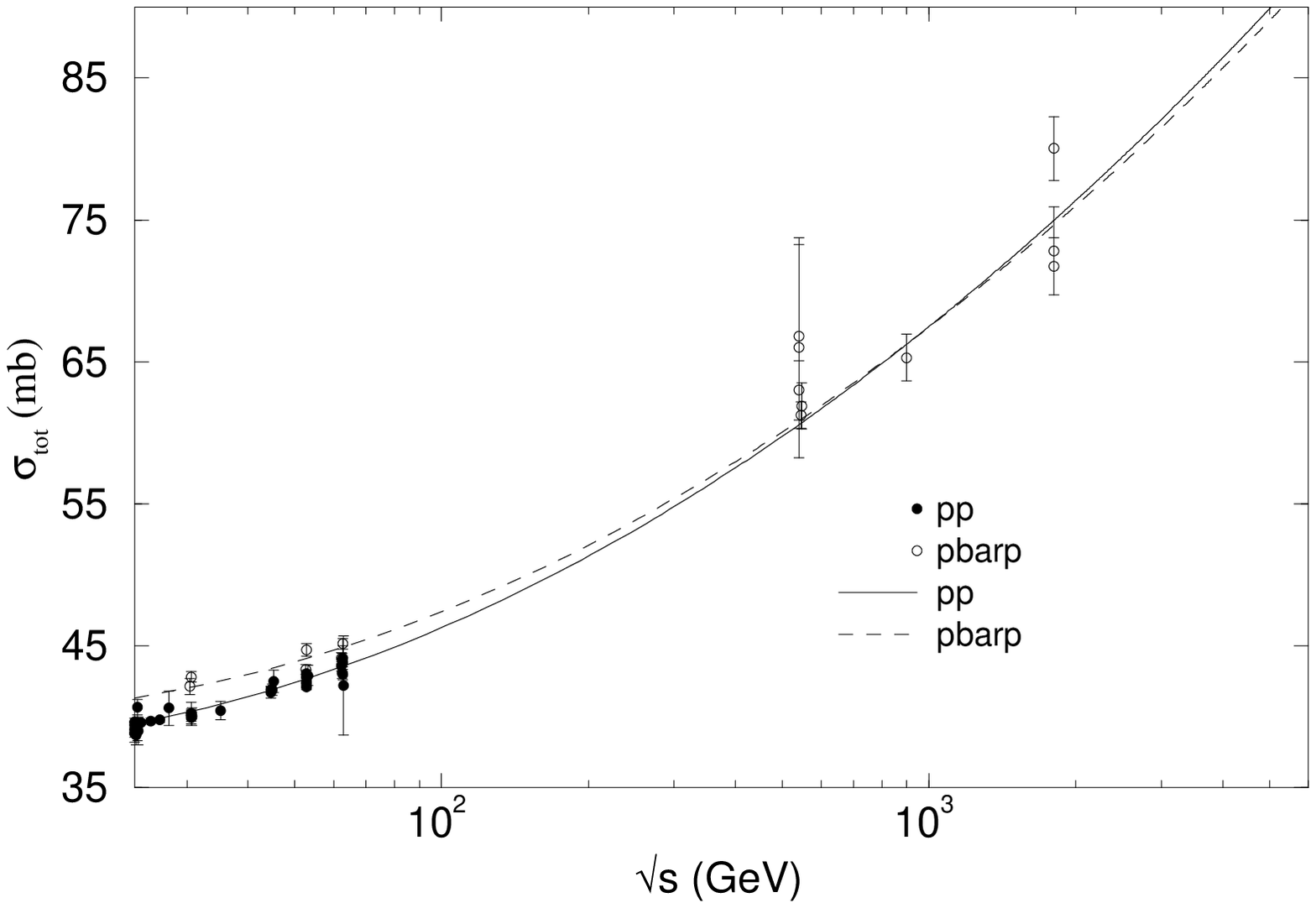}}
\resizebox{0.48\textwidth}{!}{\includegraphics{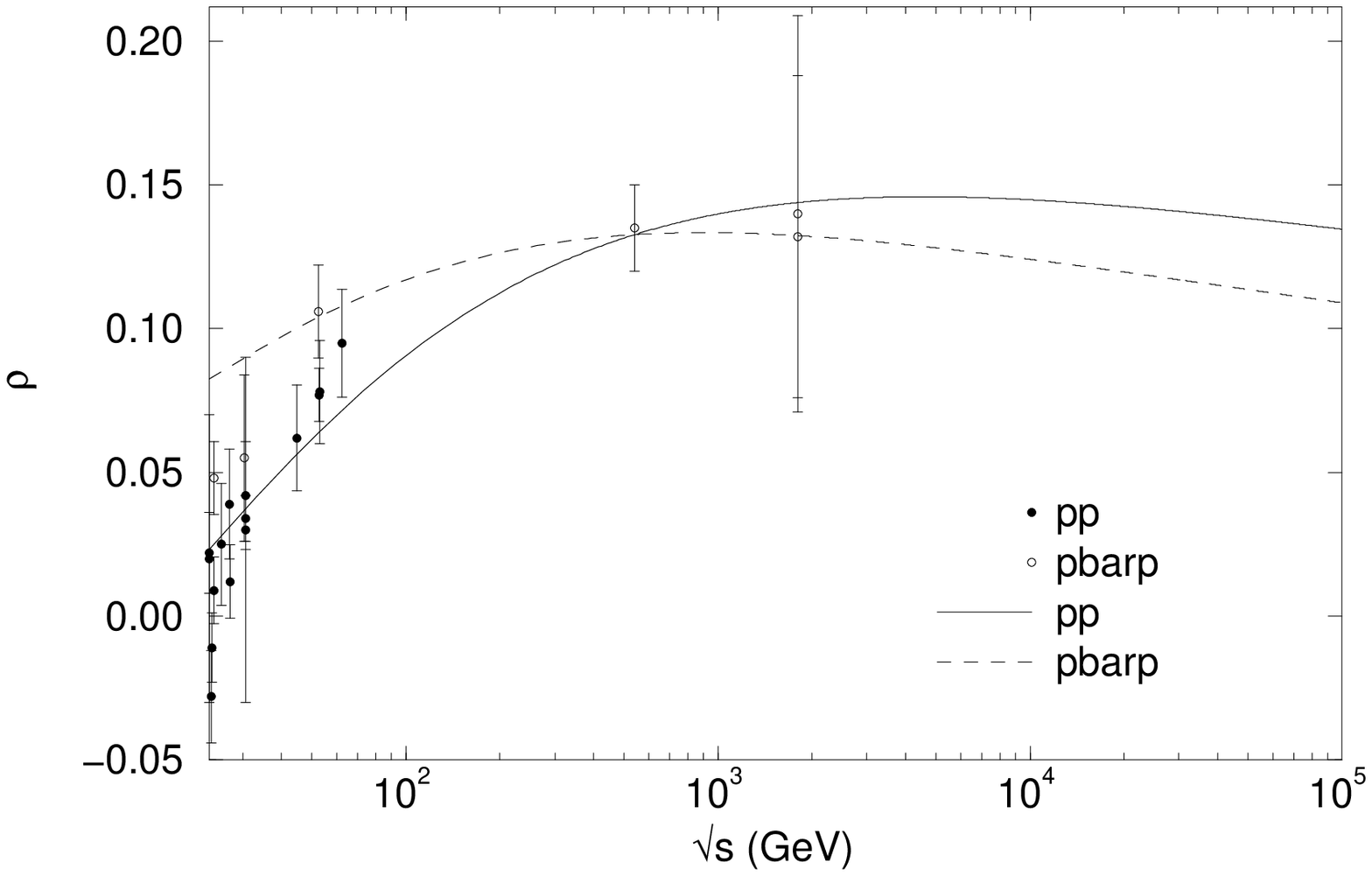}}
\caption{Total cross section and $\rho$ from global fits to $pp$ and 
$\bar{p}p$ 
data with
differential cross section data up to 
$q_{\mathrm{max}}^2$ = 2 GeV$^2$.}
\label{fig:8}       
\end{figure}

\begin{figure}
\resizebox{0.48\textwidth}{!}{\includegraphics{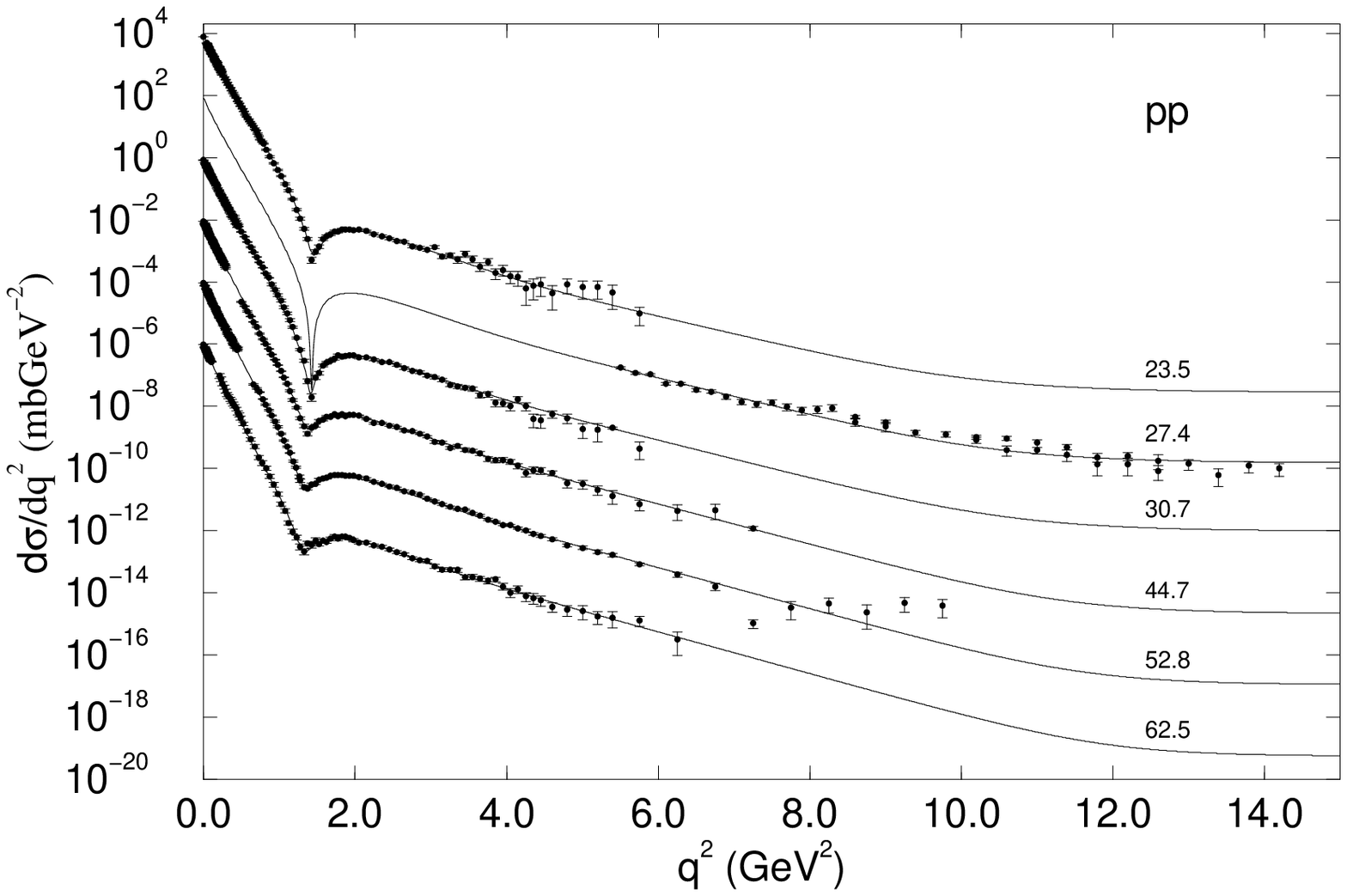}}
\resizebox{0.48\textwidth}{!}{\includegraphics{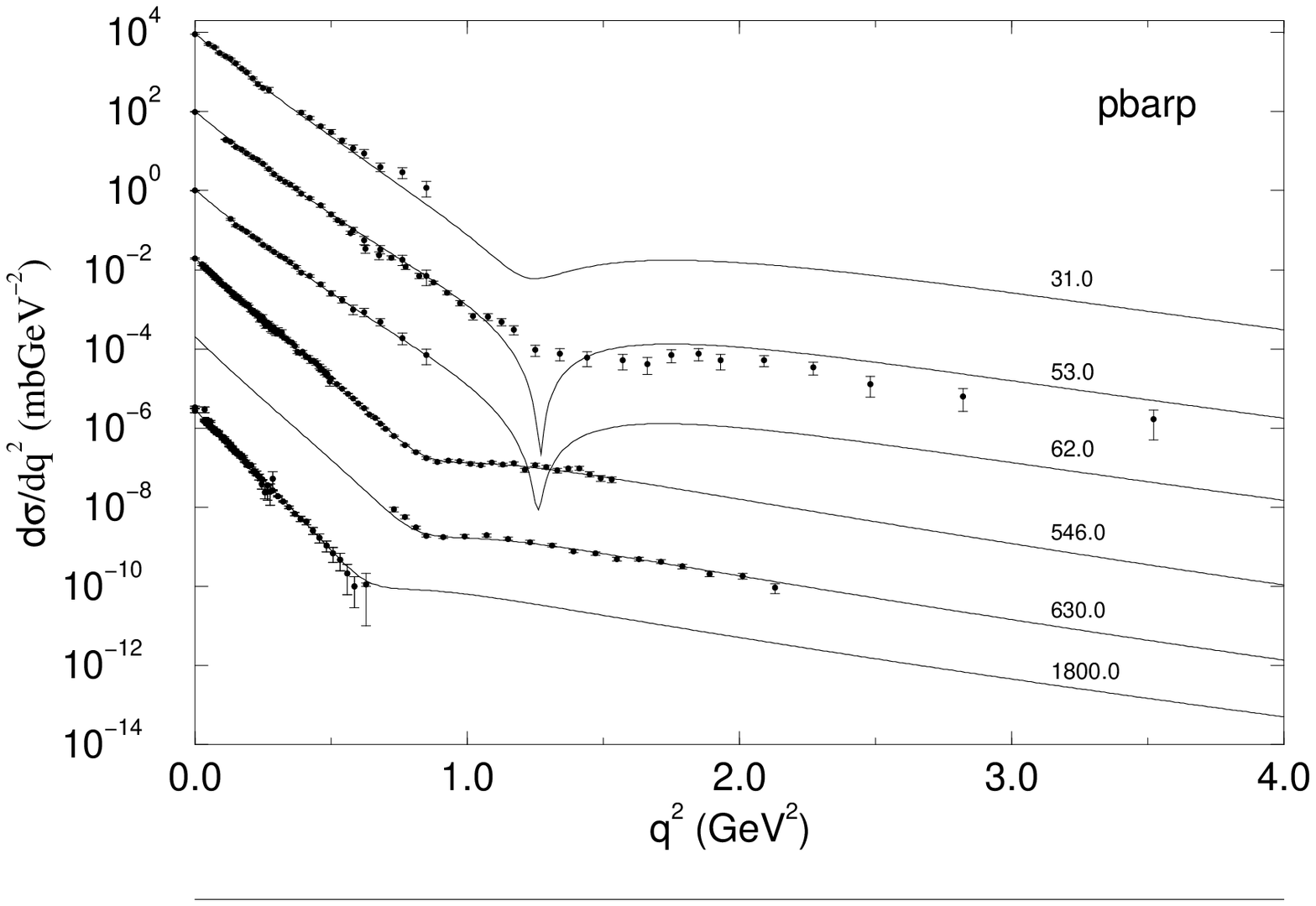}}
\caption{Differential cross sections from global fits to $pp$ and $\bar{p}p$ data with 
$q_{\mathrm{max}}^2 = 14$ GeV$^2$ (all 
differential cross section data). Curves and data were multiplied by factors
of $10^{\pm 2}$.}
\label{fig:9}       
\end{figure}

\begin{figure}
\resizebox{0.48\textwidth}{!}{\includegraphics{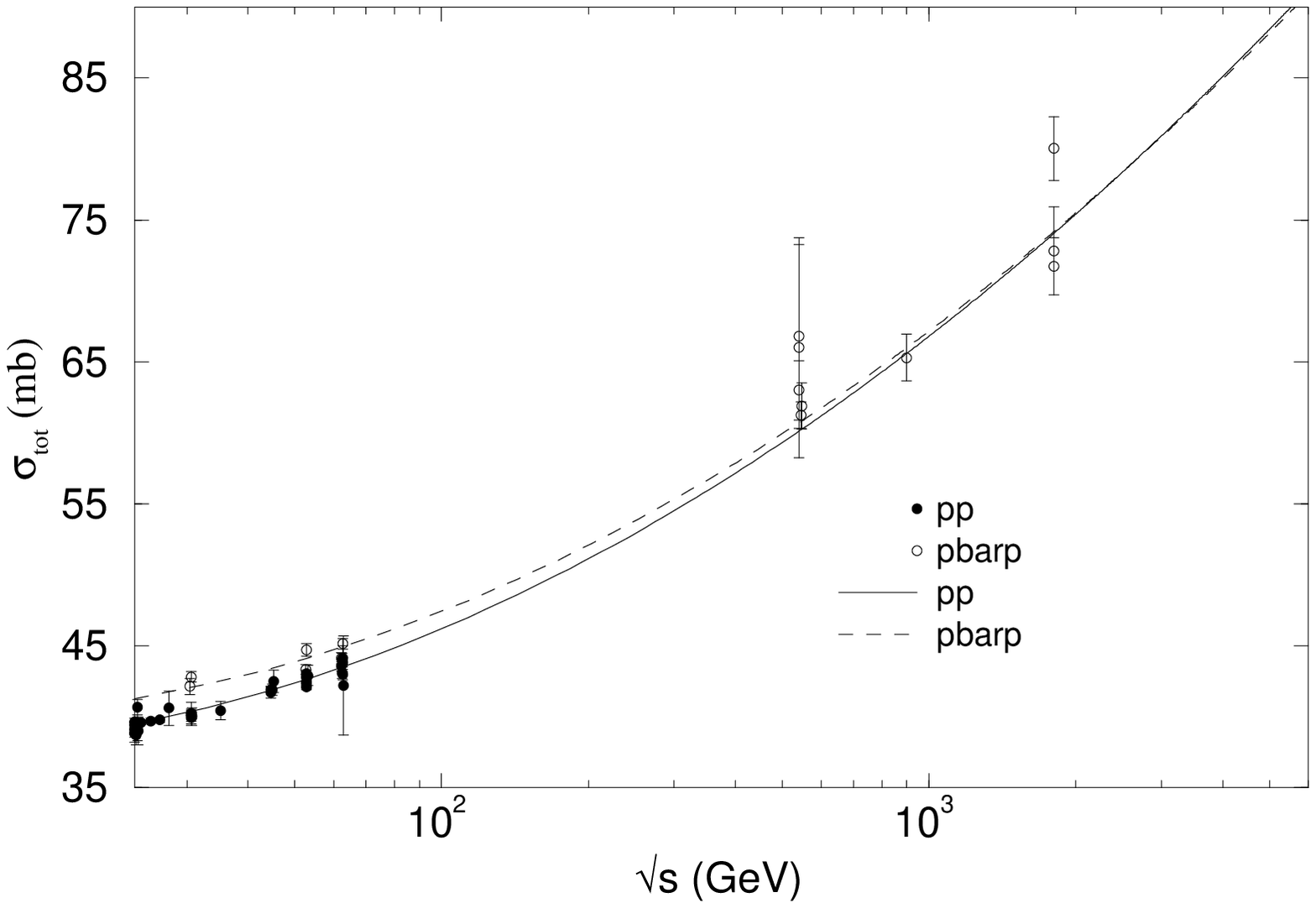}}
\resizebox{0.48\textwidth}{!}{\includegraphics{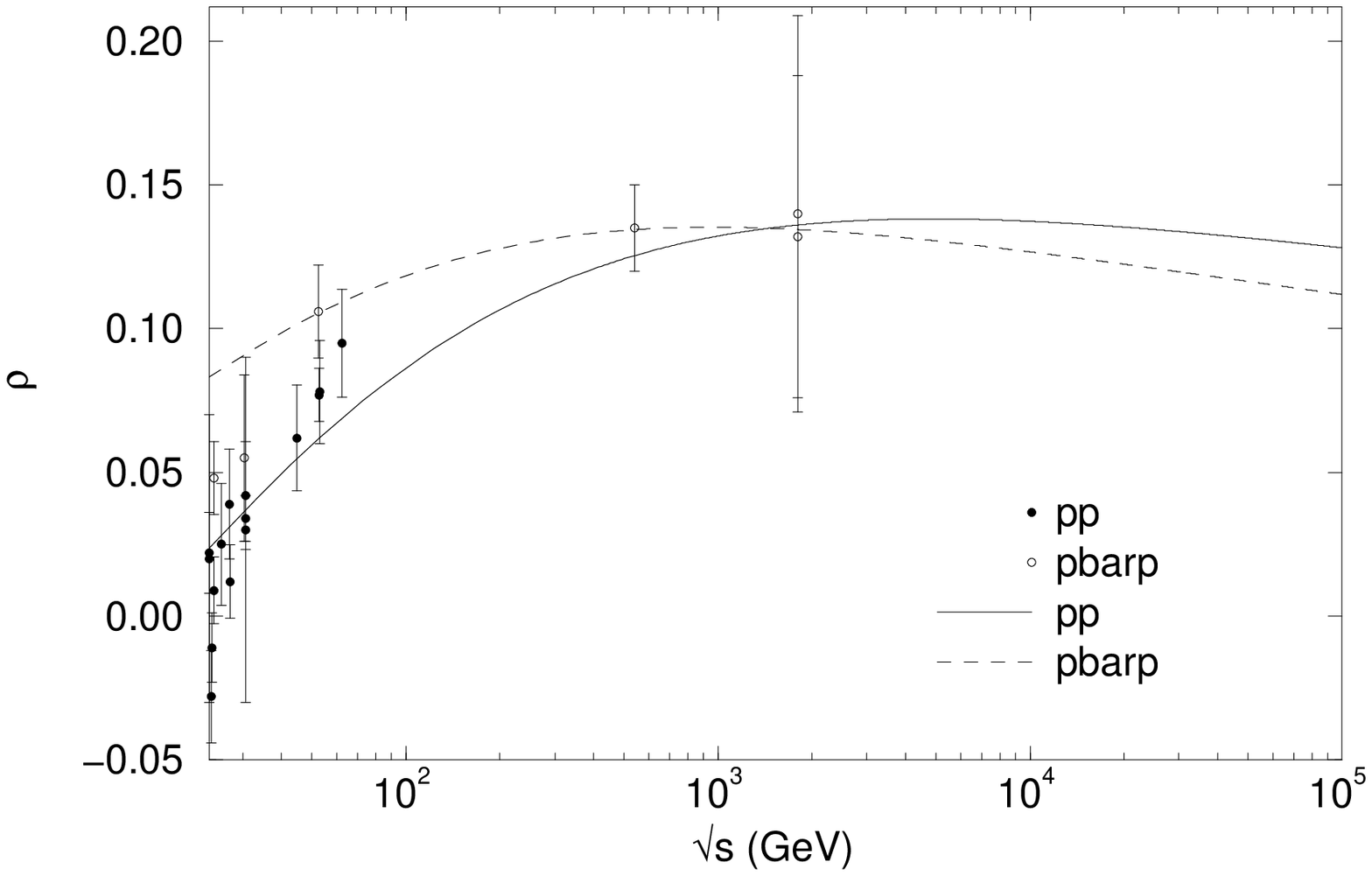}}
\caption{Total cross section and $\rho$ from global fits to $pp$ and $\bar{p}p$ 
data with
$q_{\mathrm{max}}^2 = 14$ GeV$^2$ (all the differential cross section data).}
\label{fig:10}       
\end{figure}

\section{Discussion}
\label{sec:4}

In this Section we first summarize the main results we have obtained
and then proceed with a discussion on their physical implications 
and their applicabilities in
the experimental and phenomenological contexts.

By means of a novel essentially model-independent analytical pa\-ram\-e\-trization for
the scattering amplitude and fits to physical quantities that characterize
the elastic $pp$ and $\bar{p}p$ scattering, we have developed a predictive formalism
in the variables $s$ and $q^2$, that has only empirical and formal bases. The approach
is intended for the high-energy region, specifically above $\sqrt s$ = 20 GeV
(in order to guarantee the empirical energy dependences).
We first considered global fits
to only the forward data, $\sigma_{tot}$ and $\rho$, for which dispersion
relations can be formally applied. We then included the differential cross section
data and discussed strategies for the use of the dispersion relations, namely
fits in different intervals in the momentum transfer variable:
$q_{max}^2$ = 1.0, 1.5, 2.0 and 14 GeV$^2$ (all data). The main point 
is the fact that there is no formal proof against these assumptions.
Although we have displayed here only the results for $q^2$ = 0, 
$q_{max}^2$ = 2 and $q_{max}^2$ =14 GeV$^2$, in all the cases investigated
we have obtained good descriptions of the experimental data analyzed. As commented
before, we consider the results with $q_{max}^2$ =14 GeV$^2$ as an illustrative
example on the practical applicability of the dispersion relations at medium and 
large values
of the momentum transfer. However, a striking feature is the high quality
of the data description reached in this case, as shown in Figs. (9) and (10).

In what follows, we discuss the applicability of the physical results
in the experimental and phenomenological contexts. In the former case we shall 
consider processes
that are being investigated or planned to be treated
in accelerator experiments, referring to the
following three cases: 
(1) $pp$ scattering at $\sqrt s$ = 200 GeV, that
was investigated and might yet be investigated by the pp2pp Collaboration at the
Brookhaven RHIC;
(2) $\bar{p}p$ scattering at $\sqrt s$ = 1.96 TeV, that are
being analyzed by the DZero Collaboration at the Fermilab
Tevatron (RUN II);
(3) $pp$ scattering at $\sqrt s$ = 14 TeV, planned to be investigated 
by the TOTEM Collaboration at the CERN LHC. In the phenomenological context
we shall make reference to some popular
models which are, at the same time, representatives
of different pictures to high-energy soft diffraction. To this end we will 
limit the discussion
here only to the models by Desgrolard, Giffon and Predazzi (DGP) \cite{dgp},
Bourrely, Soffer and WU (BSW) \cite{bsw}, Donnachie and Landshoff (DL)
\cite{dl}, Block, Gregores, Halzen and Pancheri (BGHP) \cite{bghp}
and the Odderon concept, introduced by Lukaszuk and
Nicolescu
\cite{odderon}.

We shall focus our discussion
on the predictions obtained for $q^2$ = 0 (forward data only), 
$q_{max}^2$ = 2 GeV$^2$ and $q_{max}^2$ = 14 GeV$^2$.
Also, we treat separately the results for the
total cross section, the $\rho$ parameter and the differential cross section,
obtained from the above three variants of the fit procedure.

\subsection{Total cross section}

In the three fit variants,
the results
indicate a crossing with $\sigma_{tot}^{pp}$ becoming greather than 
$\sigma_{tot}^{\bar{p}p}$. However, the crossing point depends on the
$q^2$-interval of the differential cross section data considered in the fit:
$\sqrt s \approx$ 100 GeV for $q^2 = 0$,
$\approx$ 500 GeV for $q_{max}^2 =$ 2 GeV$^2$ and
$\approx$ 3 TeV for $q_{max}^2$ = 14 GeV$^2$
as shown in Figs. 6, 8 and 10, respectively.
That is, asymptotically, the
difference $\Delta \sigma = \sigma_{tot}^{\bar{p}p} - \sigma_{tot}^{pp}$ 
does not goes to zero and as recalled in Sec. 2.1, this behavior
is not in disagreement with formal results obtained in the context of axiomatic
field theory.

Among the models quoted, this result suggest a dominant contribution of the
Odderon \cite{odderon} at the highest energies.
We stress that, the only information we have introduced in our parametrization
was the empirical fact that the $pp$ and $\bar{p}p$ total cross sections increase
at most as $\ln^2 s$ (parametrizations for $\alpha_i(s)$ and $\bar{\alpha}_i(s)$
in Eqs. (6) and (8)) and that the difference may increase
at most as $\ln s$ (constraint (9)).

In particular, at $\sqrt s$ = 1.80 TeV, the experimental results for 
$\sigma_{tot}^{\bar{p}p}$
are characterized by the well known discrepancies between the values
reported by the E811 and E710 Collaborations \cite{e811,e710} and
that reported by the CDF Collaboration \cite{cdf}. In this respect,
except for the forward fit result for $\sigma_{tot}^{\bar{p}p}$, which
lies between the discrepant points (Fig. 6),
the predictions including the differential cross section data
favor the E811/E710 results (Figs. 8 and 10).

In Table 4 we present our numerical predictions for the total cross sections
in the case of the experiments referred to
before. For $pp$ scattering at 14 TeV (LHC) our results with 
$q_{max}^2 = 2$ GeV$^2$ and $q_{max}^2 = 14$ GeV$^2$ are in agreement,
respectively, with the predictions from the BGHP model ($\sigma_{tot} =
108.0 \pm 3.4$ mb) \cite{bghp} and from the BSW model ($\sigma_{tot}$ =
103.5 mb) \cite{bsw}. However it should be noted that these models do
not distinguish $pp$ and $\bar{p}p$ scattering at asymptotic energies.
The Table also contains the results for $\rho(s)$, to be
discussed in what follows.

\begin{table*}
\begin{center}
\caption{Predictions for $\sigma_{tot}(s)$ and $\rho(s)$ from fits to
only forward data ($q^2 = 0$) and including the differential
cross section data ($q_{max}^2 = 2$ GeV$^2$ and $q_{max}^2 = 14$ GeV$^2$).}
\label{tab:4}
\begin{tabular}{cccccccc}
\hline\noalign{\smallskip}
\textrm{Process} &     \multicolumn{2}{c}{$q^2$ = 0}   &
\multicolumn{2}{c}{$q_{\mathrm{max}}^2$} = 2 GeV$^2$ &
\multicolumn{2}{c}{$q_{\mathrm{max}}^2$} = 14 GeV$^2$ \\
 & $\sigma_{tot}$ (mb) & $\rho$  & $\sigma_{tot}$ (mb) & $\rho$ &
$\sigma_{tot}$ (mb) & $\rho$ \\
\noalign{\smallskip}\hline\noalign{\smallskip}
$pp$ - $\sqrt s$ = 200 GeV         & 52.27 & 0.1532 & 51.32 & 0.1439 & 51.12 & 0.1065 \\
$\bar{p}p$ - $\sqrt s$ = 1.96 TeV  & 78.05 & 0.1114 & 75.74 & 0.1124 & 75.24 & 0.1343 \\
$pp$ - $\sqrt s$ = 14 TeV          & 121.6 & 0.1964 & 107.5 & 0.1321 & 105.4 & 0.1365 \\
\noalign{\smallskip}\hline
\end{tabular}
\end{center}
\end{table*}

\subsection{The $\rho$ parameter}

As a consequence of the connections between real and imaginary parts of the
amplitudes via dispersion relations, similar effects appear in our results for 
$\rho(s)$, as shown in Figs. 6, 8 and 10: $\rho^{pp}(s)$
becomes greater than $\rho^{\bar{p}p}(s)$ 
above 
$\sqrt s \approx$ 80 GeV ($q^2 = 0$),
$\approx$ 200 GeV ($q_{max}^2 = $ 2 GeV$^2$) and 
$\approx$ 2 TeV ($q_{max}^2$ = 14 GeV$^2$).
In all the cases the constraint (9) assures the asymptotic
behavior as $1/\ln s$ for both $pp$ and $\bar{p}p$ scattering.

As in the case of the total cross section,
these results are in agreement with the Odderon
dominance at the highest energies.
A crossing in $\rho(s)$ with
$\rho^{pp}(s)$
becoming greather than $\rho^{\bar{p}p}(s)$ is also
predicted in one of the versions of the DGP model \cite{dgp}
and in the analysis of Ref. \cite{alm03}, which
includes cosmic-ray information on $\sigma_{tot}^{pp}$ and a model-dependent
parametrization with Odderon contribution.

Differently from the results for the total cross sections,
we note here some distinct characteristics 
between the predictions for $\rho(s)$ obtained 
with only the forward data (Fig. 6) and those
including the differential cross section data up to $q_{max}^2$ =
2 GeV$^2$ (Fig. 8) and $q_{max}^2$ = 14 GeV$^2$ (Fig. 10).
In the former case the curve for $\rho^{\bar{p}p}(s)$ lies below the highest $\bar{p}p$ data,
what does not occur when the differential cross section data is included.
We have realized that this effect (Fig. 6 and partially
in Fig. 8) is due to the large error bars 
of the experimental data at 1.8 TeV  and also to the small number of $\rho$ data from
$\bar{p}p$ scattering above 20 GeV.
In fact, at $\sqrt s$
= 1.8 TeV, the experimental values are:
$\rho_{\mathrm{E811}} = 0.132 \pm 0.056$ \cite{e811} and
 $\rho_{\mathrm{E710}} = 0.140 \pm 0.069$ \cite{e710}, corresponding to relative
errors of 42\% and 49\%, respectively. For example, if we use the same central 
values and
reduce the errors to 10 \%, the same fit leads to a curve that pass
through the central values. However this is only a technical information that certainly
has nothing to do with a physical result in the context of our analysis. 
Experimentally it is known that,
as the energy increases, it is very difficult to reach
the Coulomb-nuclear interference region, from which the $\rho$
parameter is extracted \cite{bc}. Therefore it is not expected an improvement
in these experimental values, unless some novel technique could be developed.
In this respect, the above effect, at the highest energies, can not be eliminated
in the present formulation and fit procedure, constituting, therefore, a 
drawback in our analysis, when \textit{only forward data is considered}.
However, comparison of Figs. 6, 8 and 10 shows an interesting effect:
the quality of 
the visual description of the $\rho$ data at the highest energies 
is improved with the addition of the differential
cross section information.

The numerical predictions for $\rho(s)$, in the case of the
experiments referred to before, and from the three fit variants
($q^2$ = 0, $q_{max}^2$ = 2 GeV$^2$ and $q_{max}^2$ = 14 GeV$^2$) 
are displayed in Table 4.

\subsection{Differential cross section}

As we have shown, the descriptions of the $pp$ and $\bar{p}p$ differential cross 
section
data analyzed are quite good for $q_{max}^2$ = 2 GeV$^2$ (Fig. 7) and
even in the case of $q_{max}^2$ = 14 GeV$^2$
(Fig. 9). In this subsection we discuss the applicability of these results in the
experimental and phenomenological contexts.

To the extend that our analysis can be considered model-independent
and predictive, it may be instructive to detail the results
for the experiments referred to in the beginning of this Section.
Our predictions for these processes, from the global fit including the differential 
cross
section data up to $q_{max}^2$ = 2 GeV, are shown in Fig. 11 and the corresponding
numerical results, in the region 0 - 2 GeV$^2$, are displayed in Table 5 for some
values of the momentum transfer. From Fig. 11 we note the presence of a dip at
$q^2 \approx $  1.2 GeV$^2$ for $pp$ scattering at 200 GeV 
and that the diffraction pattern becomes a shoulder at higher energies
for both  $\bar{p}p$ ($\sqrt s$ = 1.96 TeV) 
and $pp$ scattering ($\sqrt s$ = 14 TeV); we also note the shirinkage of the
diffraction peak as the energy increases.

\begin{table*}
\begin{center}
\caption{Predictions for the differential cross sections in
mbGeV$^{-2}$ at the
RHIC, Tevatron and LHC energies, from global fits including the differential
cross section data up to $q_{max}^2$ = 2.0 GeV$^2$.}
\label{tab:5}
\begin{tabular}{llll}
\hline\noalign{\smallskip}
$q^2$ GeV$^2$ & 
$pp-\sqrt{s}=200$ $\mathrm{GeV}$& 
$\bar{p}p-\sqrt{s}=1.96$ $\mathrm{TeV}$
& $pp-\sqrt{s}=14$ $\mathrm{TeV}$  \\
\noalign{\smallskip}\hline\noalign{\smallskip}
0.00 &  136.24     & 298.18     &  602.87 \\
0.01 &  117.66     & 251.98     & 470.45 \\
0.05 &   66.649    & 129.03     & 190.54 \\
0.10 &   33.576     & 56.399     &  65.903 \\
0.15 &   17.144     & 24.839     &  23.222 \\
0.20 &   8.8167     & 10.982     &  8.2384 \\
0.25 &   4.5585     & 4.8519     &  2.9339 \\
0.30 &   2.3699     & 2.1307     & 1.0537   \\
0.35 &   1.2400     & 0.92503    &  0.39265 \\
0.40 &   0.65358    & 0.39539    &  0.16557 \\
0.45 &   0.34731    & 0.16687    &  0.090879 \\
0.50 &   0.18607    & 0.071284   & 0.067705 \\
0.55 &   0.10042    & 0.033293   &  0.060450 \\
0.60 &   0.054478   & 0.019393   &  0.057098 \\
0.65 &   0.029605   & 0.014952   &  0.054020 \\
0.70 &   0.016034   & 0.013790   &  0.050435 \\
0.75 &   0.0085958  & 0.013451   &  0.046423 \\
0.80 &   0.0045185  & 0.013078   &  0.042242 \\
0.85 &   0.0022989  & 0.012456  &  0.038117 \\
0.90 &   0.0011103  & 0.011606   &  0.034201 \\
0.95 &   0.00049312 & 0.010614   &  0.030578 \\
1.00 &   0.00018982 & 0.0095647  & 0.027284 \\
1.05 &   5.5411E-05 & 0.0085216  &  0.024324 \\
1.10 &   8.5222E-06 & 0.0075277  &  0.021686 \\
1.15 &   3.9878E-06 & 0.0066078  &  0.019345 \\
1.20 &   1.7294E-05 & 0.0057734  &  0.017276 \\
1.25 &   3.5677E-05 & 0.0050275  &  0.015449 \\
1.30 &   5.2974E-05 & 0.0043677  &  0.013838 \\
1.35 &   6.6641E-05 & 0.0037883  &  0.012417 \\
1.40 &   7.6041E-05 & 0.0032825  &  0.011164 \\
1.45 &   8.1462E-05 & 0.0028425  &  0.010058 \\
1.50 &   8.3576E-05 & 0.0024608  &  0.0090810 \\
1.55 &   8.3151E-05 & 0.0021304  &  0.0082164 \\
1.60 &   8.0907E-05 & 0.0018448  &  0.0074507 \\
1.65 &   7.7458E-05 & 0.0015981  &  0.0067714 \\
1.70 &   7.3291E-05 & 0.0013851  &  0.0061679 \\
1.75 &   6.8779E-05 & 0.0012013  &  0.0056310 \\
1.80 &   6.4189E-05 & 0.0010426  &  0.0051523 \\
1.85 &   5.9707E-05 & 0.00090562 &  0.0047250 \\
1.90 &   5.5453E-05 & 0.00078734 &  0.0043428 \\
1.95 &   5.1497E-05 & 0.00068517 &  0.0040002 \\
2.00 &   4.7875E-05 & 0.00059687 &  0.0036928 \\
\noalign{\smallskip}\hline
\end{tabular}
\end{center}
\end{table*}

\begin{figure}
\resizebox{0.48\textwidth}{!}{\includegraphics{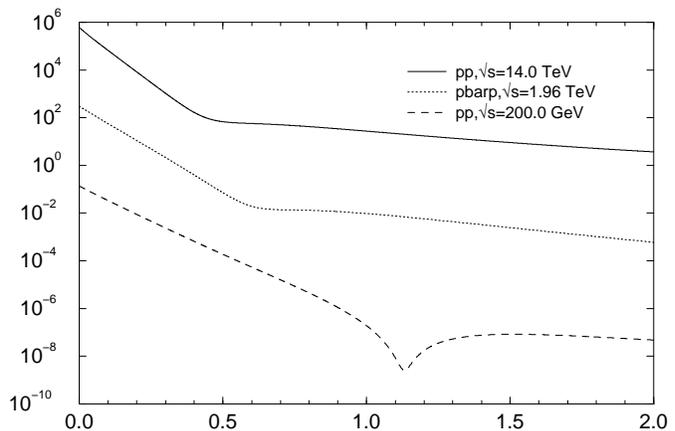}}
\caption{Predictions for the differential cross sections at the
RHIC, Tevatron and LHC energies from fits including the differential
cross section data up to $q_{max}^2$ = 2 GeV$^2$ (Table 5). The upper and lower 
curves
were multiplied by $10^3$ and $10^{-3}$, respectively.}
\label{fig:11}       
\end{figure}

\subsubsection{Experimental aspects}

As examples of practical use of these predictions in the experimental context, 
let us discuss the recent determinations of the slope parameter from elastic
rates measured by the pp2pp Collaboration ($pp$ scattering at $\sqrt s$ = 200 GeV)
at the RHIC
\cite{slopepp} and the preliminary results obtained by the DZero Collaboration
($\bar{p}p$ scattering at $\sqrt s$ = 1.96 TeV) at the Tevatron \cite{fermilab}. 
Operationally, the differential cross section is expressed by

\begin{eqnarray}
\frac{d\sigma}{dq^2} = \frac{1}{\mathcal{L}}\frac{dN}{dq^2},
\nonumber
\end{eqnarray}
where $dN/dq^2$ is the rate of the elastic interactions
and $\mathcal{L}$ the machine luminosity. Due to uncertainties
in the determination of $\mathcal{L}$ the above quoted experiments have
extracted only the slope of the elastic rates, that is, the corresponding
differential cross section could not yet be determined. In what follows we
present our results for the corresponding slopes and discuss ways to contribute
with a possible reasonable normalization of the elastic rates.

\vspace{0.5cm}

$\bullet$ \textit{$pp$ at $\sqrt s =$ 200 GeV}

In the case of the pp2pp experiment, the slope $B$ was obtained from the elastic
rates measured in the $q^2$ range
0.010 $\leq q^2 \leq$ 0.019 GeV$^2$. The corresponding amplitude has
contributions from the Coulomb amplitude, nuclear amplitude and
the interference between them, and it is  parametrized by \cite{slopepp}

\begin{eqnarray}
\frac{d\sigma}{dt} &=& 4\pi(\hbar c)^2 
\left(\frac{\alpha G^2_E}{t}\right)^2 +
\frac{1 + \rho^2}{16\pi (\hbar c)^2} \sigma_{tot}^2 e^{-B|t|} 
\nonumber \\
&-& (\rho + \Delta \Phi) \frac{\alpha G^2_E}{|t|} \sigma_{tot} 
e^{-\frac{1}{2}B|t|}. \nonumber
\end{eqnarray}

The fit parameters are the slope $B$ and a normalization constant
(elastic rates).
The input values for $\sigma_{tot}$ and $\rho$ used by the authors were
51.6 mb (obtained from the Donnachie-Landshoff model) and 0.13
(fit by the UA4/2 Collaboration), respectively. The resulting slope
parameter was 

\begin{eqnarray}
B = 16.3 \pm 1.6 \ \mathrm{(stat.)} \pm 0.9 \ \mathrm{(syst.)} \ \mathrm{GeV}^{-2}
\nonumber
\end{eqnarray}
Adding in quadratures the error reads $\pm$ 1.8 GeV$^{-2}$.

From Fig. 4 we can see that this experimental value of the slope is above the
general trend of the others measurements, even from
$\bar{p}p$ scattering. This effect is due to the small
values of the momentum transfer in which the measurement has been performed,
namely lower than those in the other experiments and also because the interval
is in the limit of the
Coulomb-nuclear interference region ($q^2 \approx $ 0.01 GeV$^2$).
Since the $pp$ data we have analyzed cover the region only up to
$\sqrt s$ = 62.5 GeV and above $q^2 =$ 0.01 GeV$^2$ (except for the
optical point), it is an important test to check our predictions for
the above quantity. 

To this end, from the fit with $q_{max}^2$ = 2 GeV$^2$ and based on the experimental 
procedure
\cite{slopepp}, we have generated 19 differential cross section points,
with estimated error of 1\%, in the region 0.010 $ \leq q^2 \leq $ 0.019
GeV$^2$ and fitted the points with an exponential form in the momentum transfer 

\begin{eqnarray}
\frac{d\sigma}{dq^2} = A e^{-Bq^2},
\end{eqnarray}
as shown in Fig. 12. With this procedure we have obtained 

\begin{eqnarray}
A &=& 136.0 \pm 1.7 \ \mathrm{mbGeV}^{-2}, \nonumber \\
B &=& 14.46 \pm 0.84 \ \mathrm{GeV}^{-2}, \nonumber
\end{eqnarray}
with $\chi^2$/DOF = $4.8 \times 10^{-5}$ for 17 degrees of freedom.
Therefore, our result for the slope is in agreement with the experimental value,
lying inside the lower error bar in the case that
statistical and systematic errors are added in quadrature. The relative
error in respect the central value is  11\%.
Moreover, the input value  used by the pp2pp Collaboration,
$\sigma_{tot}^{pp}$ = 51.6 mb, is also in agreement with our predictions for
the cross section, as shown in Table 4, namely 51.32 mb. We also note that, although our results
indicate $\rho$ = 0.1439, which is higher than the input $\rho$ = 0.13,
this difference has no practical effect in the nuclear contribution in Eq. (16), since
this parameter appears in the form $1 + \rho^2$. However that is not the case for the
total cross section which has a quadratic contribution: $\sigma_{tot}^2$. We understand 
that these
results corroborates the accuracy of our predictions and in this sense, the
above value we obtained for the parameter $A$ could be used as a suitable
normalization factor in the estimation of the corresponding
differential cross section.

\begin{figure}
\resizebox{0.48\textwidth}{!}{\includegraphics{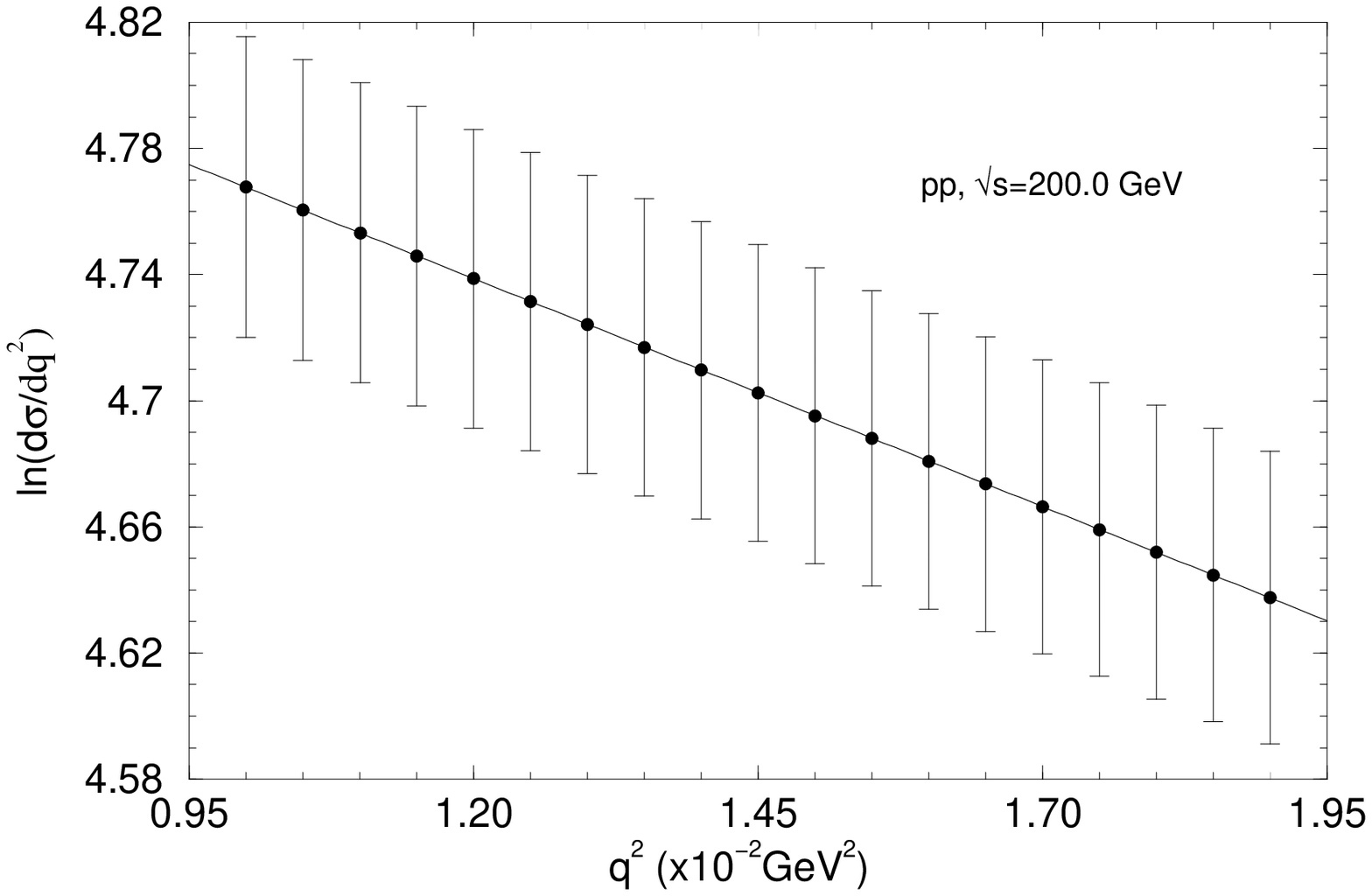}}
\resizebox{0.48\textwidth}{!}{\includegraphics{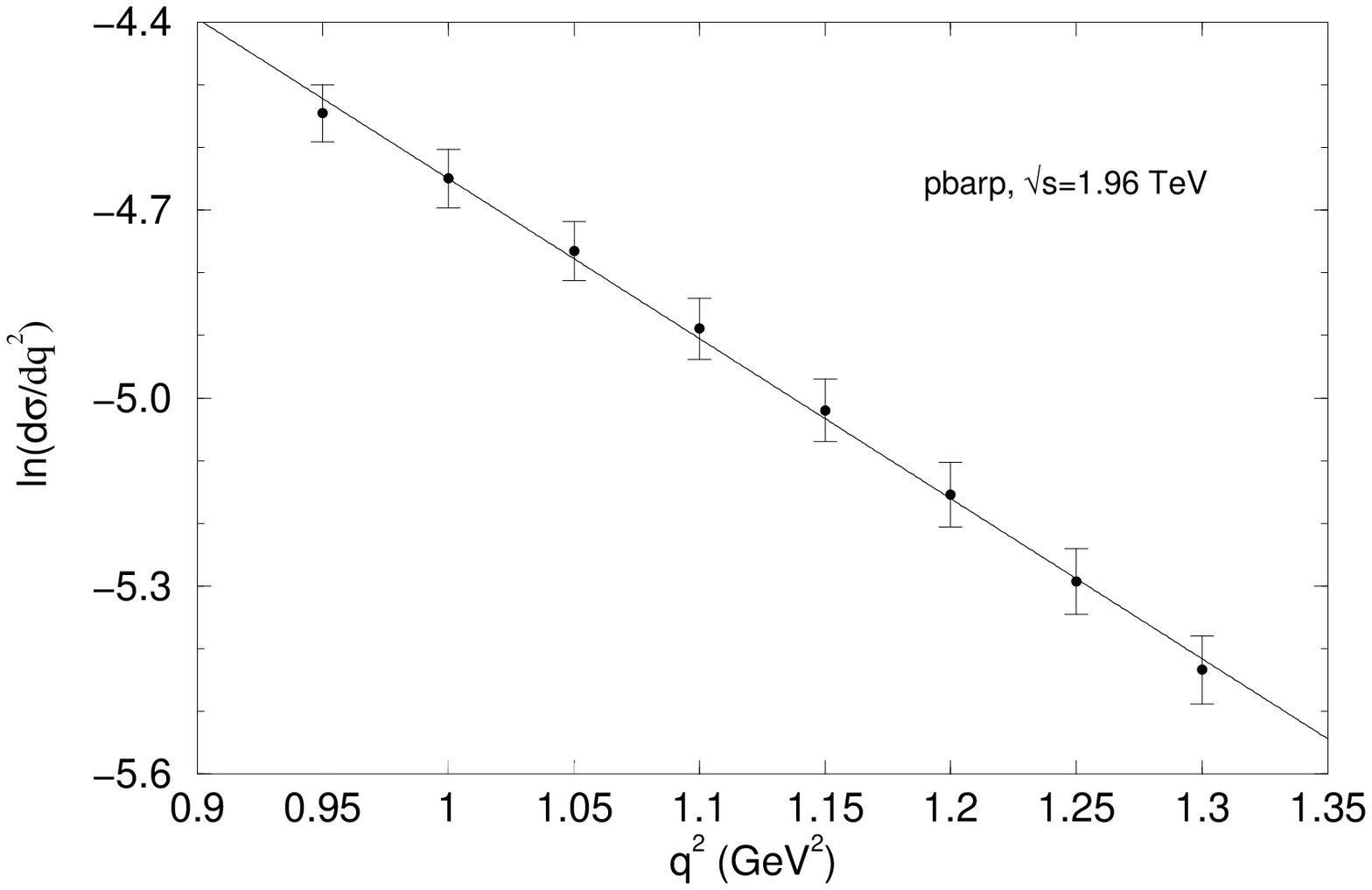}}
\caption{Determination of the slope by means of an exponential
fit, Eq. (16), to the generated differential
cross section points, for $q_{max}^2 =$ 2 GeV$^2$.}
\label{fig:12}       
\end{figure}

\vspace{0.5cm}

$\bullet$ \textit{$\bar{p}p$ at $\sqrt s =$ 1.96 TeV}

Now let us discuss the recent 
measurements (even if preliminary) of the elastic rates performed by the DZero 
Collaboration,
from $\bar{p}p$ scattering at $\sqrt s$ = 1.96 TeV \cite{fermilab}. In this case, 
the rate
of elastic collisions has been
measured at medium values of the momentum transfer, 
in the interval 0.96 $< q^2 <$ 1.31 GeV$^2$ \cite{molina}. 
As an illustration, and for further discussion,
our prediction 
for the differential cross section at $\sqrt s =$ 1.96 TeV and 
$\sqrt s =$ 1.80 TeV, with $q_{max}^2 = 2$ GeV$^2$, are shown in Fig. 13
together with the experimental data, obtained by the E710 and CDF Collaborations,
at $\sqrt s =$ 1.80 GeV.

\begin{figure}
\resizebox{0.48\textwidth}{!}{\includegraphics{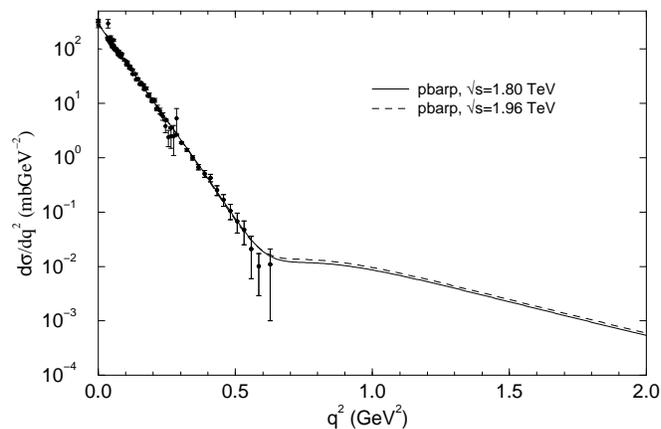}}
\caption{Prediction for the $\bar{p}p$ differential cross section at 
$\sqrt s =$ 1.96 TeV
and  $\sqrt s =$ 1.80 TeV, for $q_{max}^2$ = 2 GeV$^2$, together with
and experimental data at  $\sqrt s =$ 1.80 TeV.}
\label{fig:13}       
\end{figure}

In principle, the elastic rates at 1.96 TeV could be compared with
the differential cross section data at 1.80 TeV, allowing a kind of
normalization.
However, from Fig. 13, we see that the E710 data cover the
region only up to $q^2$ = 0.627 GeV$^2$ and the main problem is the fact that in the
gap between this last point and the first DZero point ($q^2 \approx $ 0.96 GeV$^2$)
it is expected the presence of a dip or a shoulder, implying, in any case, 
in a change
of curvature. What is worst, the E710 point at $q^2$ = 0.627 GeV$^2$ has
a large error bar (not shown in the figure, but
taken into account in all the fits), turning out very difficult, in our oppinion, 
any attempt to 
perform a reasonable normalization. 

In this respect,
looking for a more quantitative information and, as before, from the fit with
$q_{max}^2$ = 2 GeV$^2$ and
 based on the experimental
procedure \cite{molina}, we have generated 8 differential cross
section points with errors of 1\%, in the region 0.95 $\leq q^2 \leq$ 1.3 GeV$^2$
and fitted the points with the exponential form, as shown in Fig. 12. In this
case we have obtained

\begin{eqnarray}
A &=& 0.123 \pm 0.004 \ \mathrm{mbGeV}^{-2}, \nonumber \\
B &=& 3.554 \pm 0.030 \ \mathrm{GeV}^{-2}, \nonumber
\end{eqnarray}
with $\chi^2$/DOF = 2.52 for 6 degrees of freedom.
We note that a close looking at the generated points in Fig. 12
shows that the last four points have a slope slightly greater than
the first four points and this effect seems also to be present in
the measured elastic rates \cite{fermilab,molina}.
Although the experimental data are still being analyzed by the DZero Collaboration,
we understand that the above information and the numerical results displayed
in Table 5, can contribute with the discussion on a suitable normalization
for these elastic rates. We shall return to this point in what follows.

\subsubsection{Phenomenological aspects}

We now turn the discussion to the phenomenological context, with main focus on the
result we have obtained for the highest energy with 
differential cross section data available,
namely $\bar{p}p$ scattering at $\sqrt s$ = 1.80 TeV, Fig. 13. The point is to 
compare this result
with predictions from the models referred to in the beginning of this Section. 

From Fig. 13, our result indicates a change of curvature in the region of the
last three experimental points ($q^2 \approx 0.55 - 0.65$ GeV$^2$) with a
shoulder shape and not a dip (minimum) with defined position. This effect is
due to the contribution from the real part of the amplitude as shown in Fig. 14,
where we display separately the contributions to the differential cross section 
from only the
real and only the imaginary parts of the amplitudes in the cases of
$q_{max}^2$ = 2 GeV$^2$ and 
$q_{max}^2$ = 14 GeV$^2$.
From this Figure we see that, as expected, the imaginary part presents a
zero (change of sign) and inside this region, the value of the minimum in the
differential cross section is due to the contribution of the real part
(a shoulder in this case). The real part of the amplitude also
presents a zero at $q_{0}^2 \approx$ 0.30 GeV$^2$ in the case
of $q_{max}^2$ = 2 GeV$^2$ and
$q_{0}^2 \approx$ 0.38 GeV$^2$ for
$q_{max}^2$ = 14 GeV$^2$. These results for the real part are in agreement with
a theorem demonstrated by Martin, which states that the real part changes
sign at $q^2 > $ 0.1 GeV$^2$ \cite{teomartin}.

On the other hand, the contributions from the imaginary parts are very similar
in both cases, indicating
a zero at $q_{0}^2 \approx$ 0.70 GeV$^2$
for $q_{max}^2$ = 2 GeV$^2$ and at $q_{0}^2 \approx$ 0.73 GeV$^2$
for $q_{max}^2$ = 14 GeV$^2$. Therefore, from this Figure, we can infer with some security that the 
position of the first minimum in the differential cross section 
at this energy occurs at $q_{0}^2$ = 0.70 GeV$^2$ ($q_{max}^2$ = 2 GeV$^2$).
In the phenomenological context this value is in agreement with the
predictions of the DGP, BSW and DL models, but not with that from the BGHP model,
since the minimum in this model is predicted to be at $q_{0}^2 \approx $0.60 GeV$^2$ 
(coincident with the highest E710 point). We think that this is an important point,
that should be carefully analyzed, when comparing elastic rates with model predictions
at $\sqrt s$ = 1.96 TeV.

\begin{figure}
\resizebox{0.48\textwidth}{!}{\includegraphics{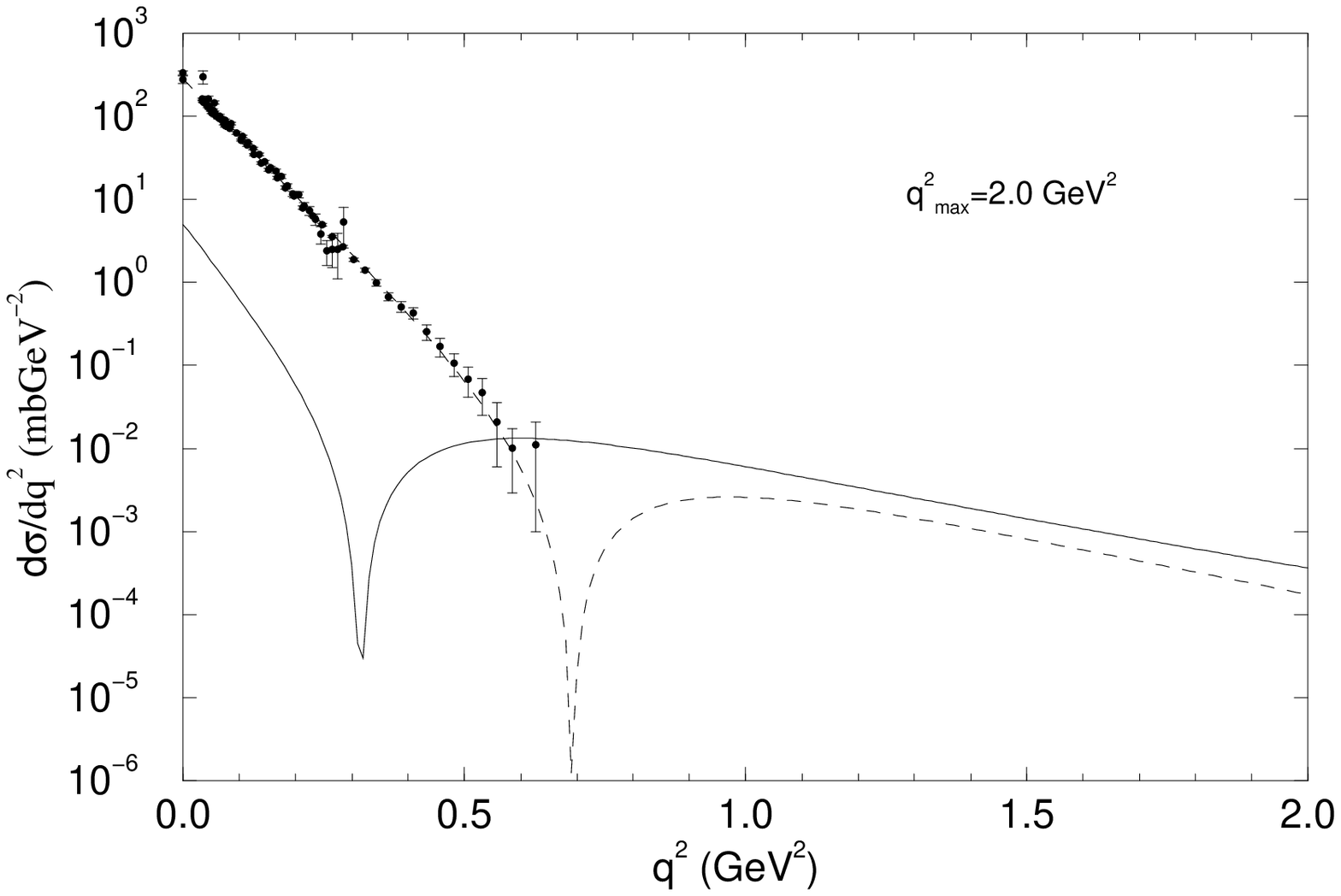}}
\resizebox{0.48\textwidth}{!}{\includegraphics{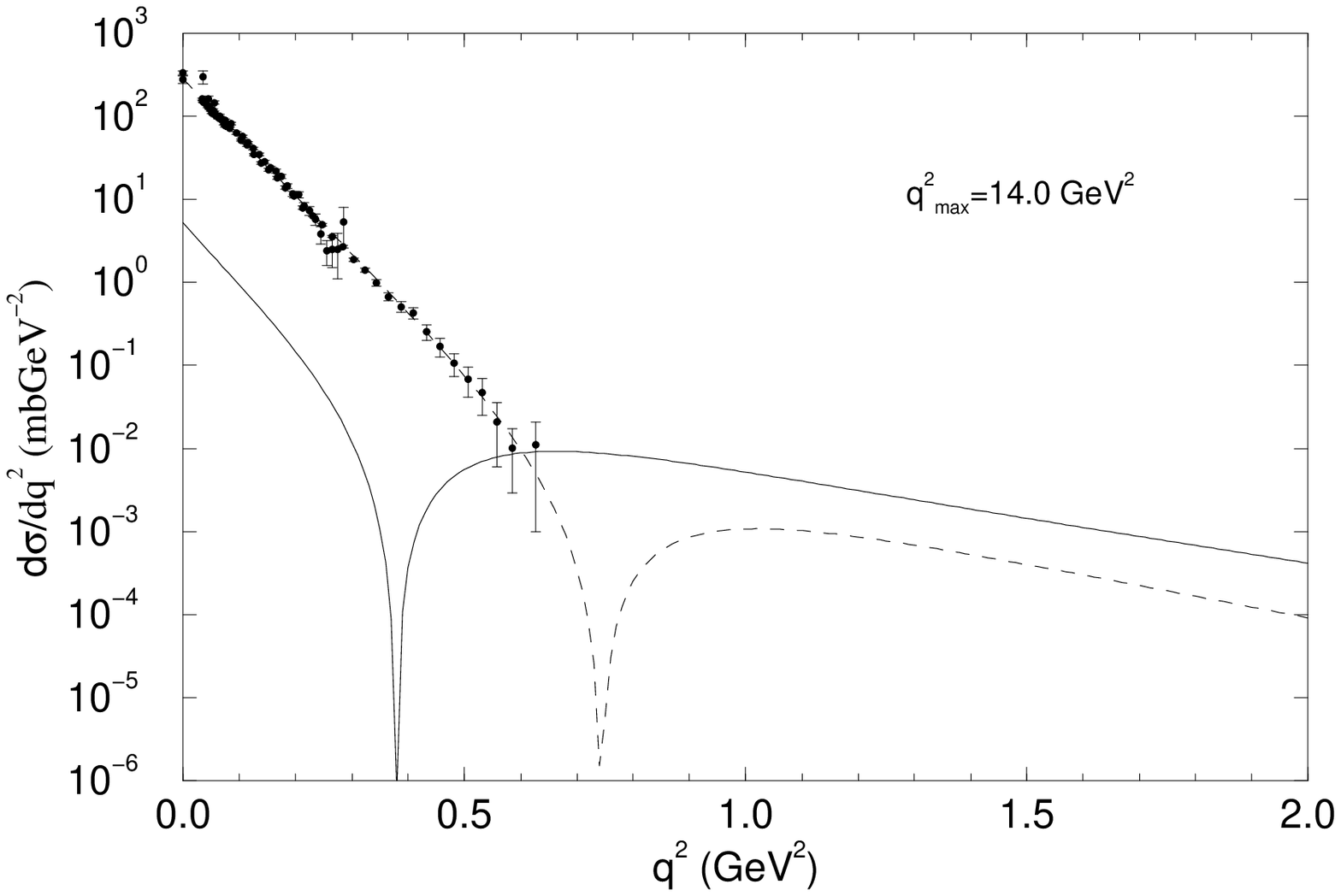}}
\caption{Contributions to the $\bar{p}p$ differential cross section 
at $\sqrt s$ = 1.80 TeV from the
real (solid) and imaginary (dashed) parts of the amplitude for $q_{max}^2$ = 2 GeV$^2$
and
$q_{max}^2$ 14 GeV$^2$.}
\label{fig:14}       
\end{figure}

Another aspect to note in Fig. 14 is that, in both cases, the contribution of the
imaginary part dominates in the region of small momentum transfer,
up to the beginning of the shoulder. On the other hand, in this region and for
higher values of the momentum transfer, it is the contribution
of the real part that dominates. However, in order to investigate this effect
in more detail,
we must consider the region of medium and large momentum transfer,
that is the results of the fits with $q_{max}^2$ = 14 GeV$^2$.
We stress that, even under restrictive formal justification,
our results taking into account all the differential cross section data
are quite good, as shown in Fig. 9 and therefore it may  be instructive
to discuss the implications of this variant of the fit.

Concerning 
the contributions to the differential cross  sections from the real and  the
imaginary parts of the amplitude, we consider three typical examples:
the
results for $pp$ scattering at 52.8 GeV and $\bar{p}p$ at 53 GeV, shown in Fig. 15, and 
those for $\bar{p}p$ at 546 GeV, displayed in Fig. 16, 
together with the corresponding experimental data. 
The point is that, according to our predictions, in the 
energy region of the CERN ISR
($\sqrt s \approx $ 23 - 63 GeV), the imaginary part dominates at 
medium and large values of the momentum transfer (Fig. 15). On the other
hand, at 
higher energies, such as the regions of the CERN  Collider (Fig. 16)
and Tevatron  (Fig. 14), it is the contribution from the real
part that dominates. 

\begin{figure}
\resizebox{0.48\textwidth}{!}{\includegraphics{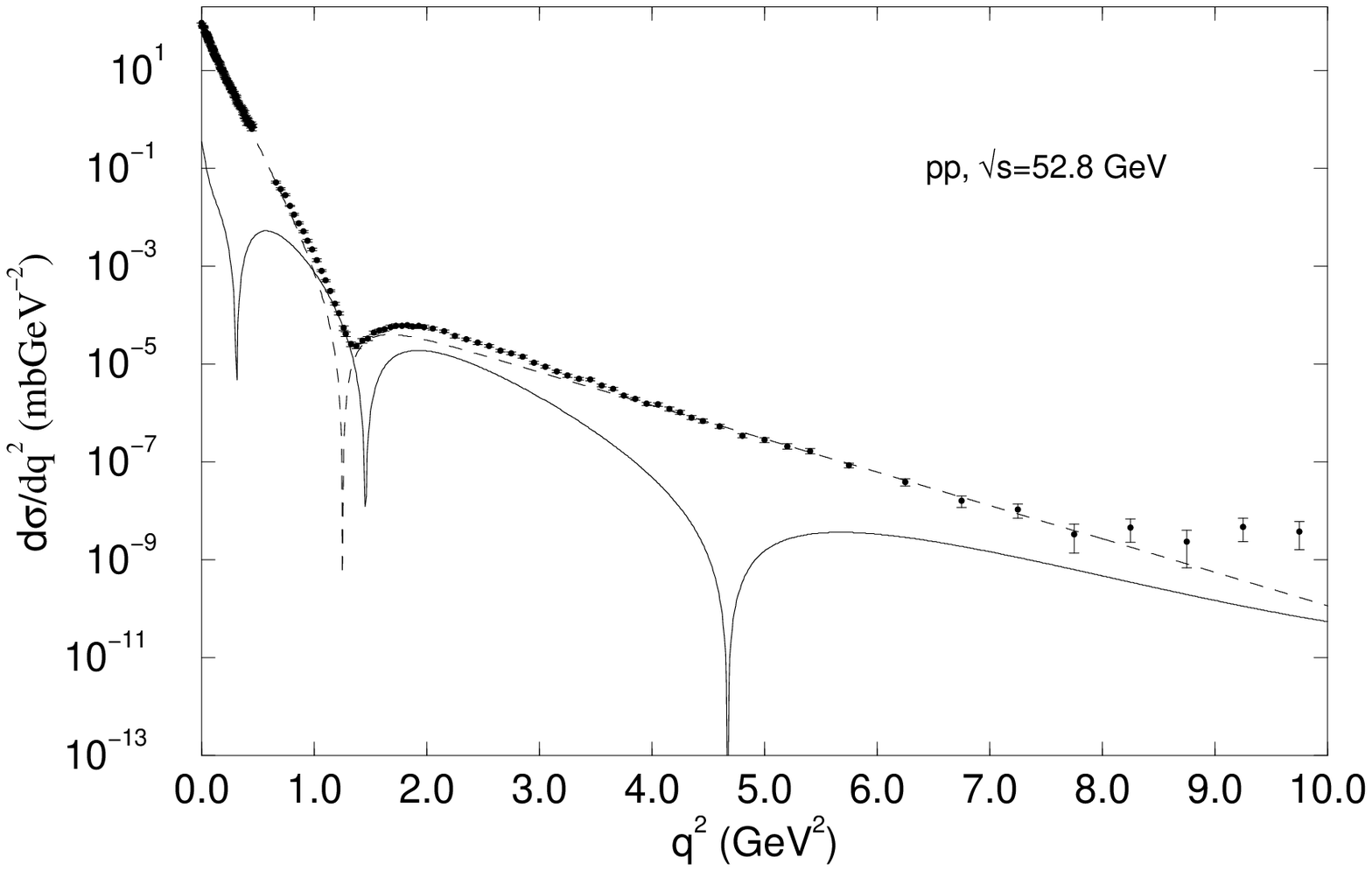}}
\resizebox{0.48\textwidth}{!}{\includegraphics{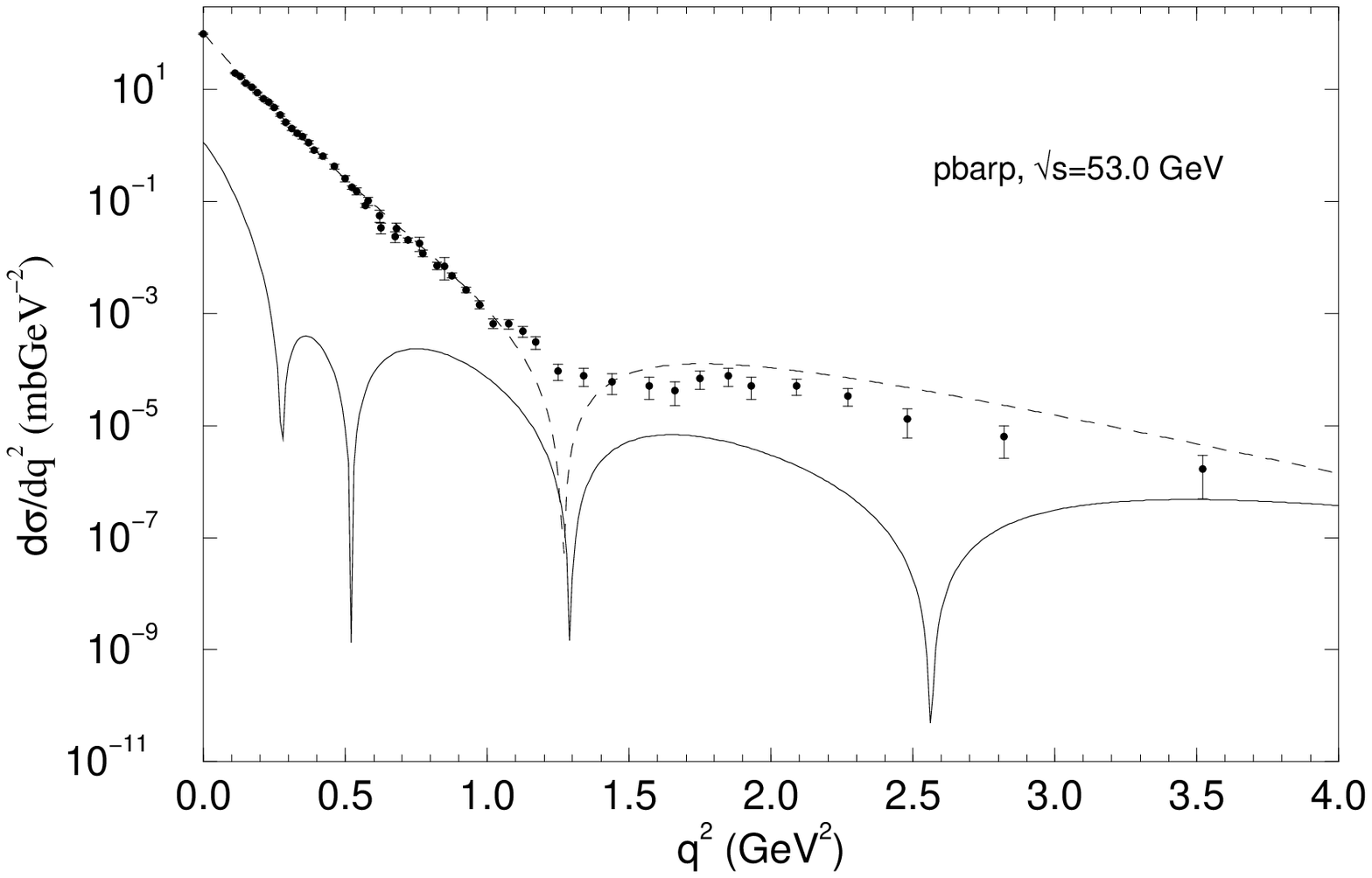}}
\caption{Contributions to the differential cross section from the
real (solid) and imaginary (dashed) parts of the amplitude for 
$q_{max}^2$ = 14 GeV$^2$.}
\label{fig:15}       
\end{figure}

\begin{figure}
\resizebox{0.48\textwidth}{!}{\includegraphics{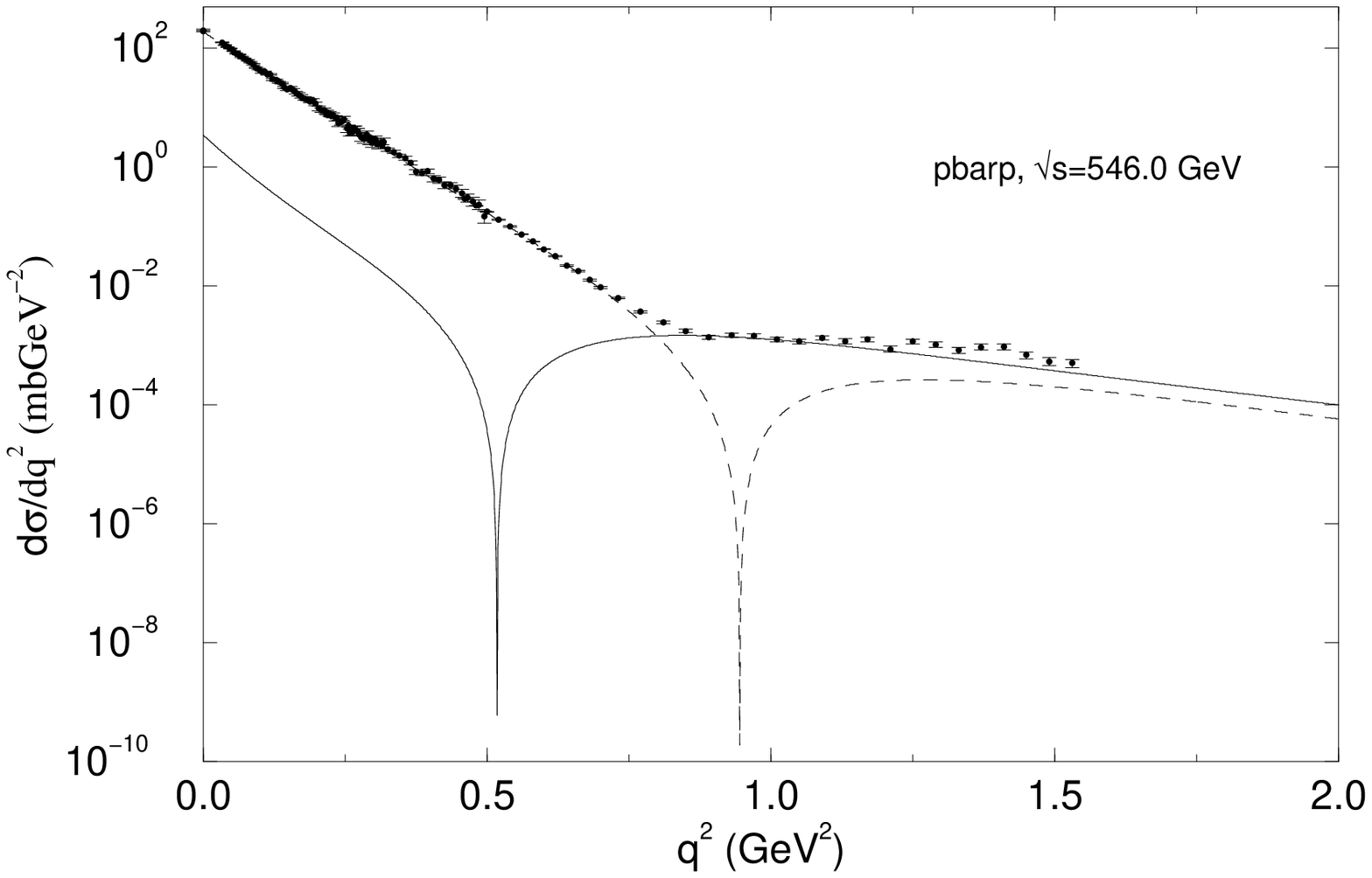}}
\caption{Contributions to the differential cross section from the
real (solid) and imaginary (dashed) parts of the amplitude for 
$q_{max}^2$ = 14 GeV$^2$.}
\label{fig:16}       
\end{figure}

At last, it may also be instructive to see what kind of results can be predicted
in the region of large momentum transfer at the
RHIC, Tevatron and LHC energies.
We display that in Fig. 17 up to $q^2$ = 8 GeV$^2$, the interval generally
considered in the publications.
The main point here is the prediction of
a smooth decrease of the differential cross section
above the first minimum,  without secondary structures in this region.
Among the quoted phenomenological approaches, this behavior is predicted only in the 
DL model. However, we note from Fig. 9 that a small change in the curvature
is predicted at $q^2 \approx $ 12 GeV$^2$.

\begin{figure}
\resizebox{0.48\textwidth}{!}{\includegraphics{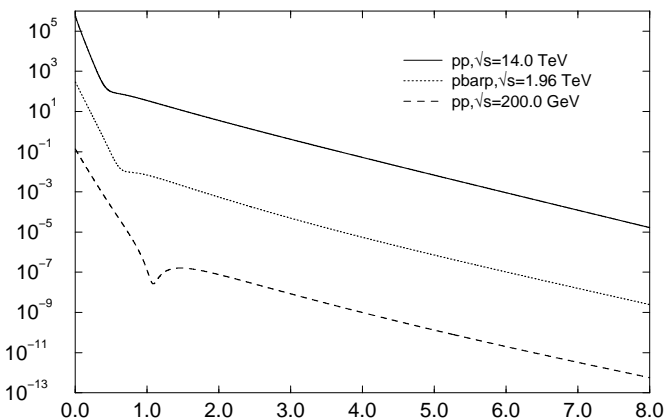}}
\caption{Predictions for the differential cross sections at the
RHIC, Tevatron and LHC energies from fits including the differential
cross section data up to $q_{max}^2$ = 14 GeV$^2$ (all data).The upper and lower 
curves
were multiplied by $10^3$ and $10^{-3}$, respectively.}
\label{fig:17}       
\end{figure}

\section{Conclusions and Final Remarks}
\label{sec:5}

We have introduced an analytical parametrization for the
elastic hadron-hadron scattering amplitude and a fit
procedure characterized by at least five important novel aspects:
(1) the parametrization is almost model-independent, with enclosed
dependences on the energy and momentum extracted from the empirical behavior 
of the experimental data and in agreement with some high-energy theorems
and bounds from AQFT;
(2) the real and imaginary parts of the amplitude
are entire functions of the logarithm of the energy $s$ and are connected
through derivative dispersion relations; (3) the $pp$ and $\bar{p}p$
scattering are also connected to the extent that analyticity 
and unitarity lead to
dispersion relations; (4) the approach is predictive in both energy 
and momentum variables;
(5) fits to $pp$ and $\bar{p}p$ experimental data, above 20 GeV, on the 
forward quantities
and then including differential cross sections,
allow good global descriptions of all the data, even in different 
regions of the momentum transfer.

We have presented a critical remark on a drawback that still remains in the
present formulation, which is related to the results for $\rho^{\bar{p}p}(s)$, 
in the particular case of forward fit. One way to address this question may be 
to consider the derivative dispersion relations up to second or third order
in the tangent operator. Results on this direction will be reported elsewhere.

Another aspect that deserves some comment is the number of free parameters
involved in the analysis.
When including the differential cross section data,
even in the region $q^2 \leq$ 1 GeV$^2$, the fit demands 3 exponentials in the
imaginary part of the amplitude and therefore 30 free fit parameters. We understand
that this can not be seen as an disadvantage of the formalism in terms of a 
large number of 
parameters. In fact,
we are not treating with a theoretical model but, on the contrary, a model-independent 
approach aimed to describe and predict the physical quantities 
of interest on empirical and formal grounds. Therefore, the number of parameters 
does not
matter and, in this context, it can be as large as it is needed.

In this analysis we made use of the standard sets of experimental data 
on $pp$ and $\bar{p}p$ scattering above 20 GeV
(referred to in Sect. 3.1), without any kind of data
selection. As commented, this strategy explains the large values of the
$\chi^2$/DOF in the fits. However it is important to mention that
recent analyses point out the necessity of some screening criterion
in order to avoid spurious data, normalization problems and other effects in both
forward and nonforward data 
\cite{block,clm}. All that could improve the quality of the fits and will be subject of future 
investigation.

We also note that we did not use any model information in the construction
of parametrizations (5-8): they were inferred only with basis on the empirical
behavior of the experimental data above 20 GeV. However, from a phenomenological
point of view, it is expected that some contributions from lower energies
may still be present at the above threshold (for example, secondary mesonic
exchanges, in a Regge context \cite{cmsl04}). Therefore it may be interesting
to test additional terms in our original parametrization, that could simulate
these effects, from an empirical point of view, and investigate the consequences
in the description of the experimental data.

We now summarize some results we understand are topical
in this analysis. The behavior of the forward quantities,
$\sigma_{tot}(s)$ and $\rho(s)$, from $pp$ and $\bar{p}p$ scattering
are characterized by crossing effects, which are typical of Odderon
contributions. A relevant result is the prediction that $\rho^{pp}$ becomes
higher than  $\rho^{\bar{p}p}$ above $\sqrt s \approx$ 80 GeV, a result
that might be verified in short term at the RHIC by the pp2pp Collaboration.
 Our results for the differential cross section at the
Tevatron energies are in agreement with the predictions from the
majority of the models, except that by Block et al. \cite{bghp},
in what concerns the position of the first minimum.
We have also discussed the applicability of our numerical results in the normalization
of elastic rates. We add that, if we consider
the fit including all the differential cross section data, the DL model
is favored, since no structures is predicted in the region of large 
momentum transfer. 

In closing we should stress that, despite the encouraging results we have 
reached,
this phenomenological analysis constitutes a first
attempt in the search of a formally rigorous and predictive model-independent
approach. Much more research must still be done along several lines, as for
example, a complete check on all the high-energy theorems and bounds,
to establish
the exact interval in the momentum transfer variable in which dispersion
relations hold (or another framework for evaluation of the real part
of the amplitude), studies on the effect of higher orders in the derivative dispersion
relations and a systematic investigation on the influence of data selection.
We hope that the results here presented can contribute with further developments
along these aspects.

\begin{acknowledgement}
For financial support R.F.A. and M.J.M.  are thankful to FAPESP  
(Contracts No. 03/00228-0 and No. 04/10619-9) and S.D.C. to BIG - UNICAMP.
We are grateful to J. Molina and E.G.S. Luna for discussions.
\end{acknowledgement}

\end{document}